\documentclass[twocolumn]{aastex631} 

\usepackage{multirow}
\usepackage{amsmath}
\usepackage{apjfonts} 
\usepackage{comment}
\usepackage{enumitem}

\received{xx xx xxxx}
\revised{xx xx xxxx}
\accepted{xx xx xxxx}

\shorttitle{Ronayne et al.}
\shortauthors{Ronayne et al.}

\begin{document}
\title{CEERS: 7.7 \(\mu m\) PAH Star Formation Rate Calibration with JWST MIRI}

\author[0000-0001-5749-5452]{Kaila Ronayne}
\affiliation{Department of Physics and Astronomy, Texas A\&M University, College Station, TX, 77843-4242 USA}
\affiliation{George P.\ and Cynthia Woods Mitchell Institute for Fundamental Physics and Astronomy, Texas A\&M University, College Station, TX, 77843-4242 USA}

\author[0000-0001-7503-8482]{Casey Papovich}
\affiliation{Department of Physics and Astronomy, Texas A\&M University, College Station, TX, 77843-4242 USA}
\affiliation{George P.\ and Cynthia Woods Mitchell Institute for Fundamental Physics and Astronomy, Texas A\&M University, College Station, TX, 77843-4242 USA}

\author[0000-0001-8835-7722]{Guang Yang}
\affiliation{Kapteyn Astronomical Institute, University of Groningen, P.O. Box 800, 9700 AV Groningen, The Netherlands}
\affiliation{SRON Netherlands Institute for Space Research, Postbus 800, 9700 AV Groningen, The Netherlands}

\author[0000-0001-9495-7759]{Lu Shen}
\affiliation{Department of Physics and Astronomy, Texas A\&M University, College Station, TX, 77843-4242 USA}
\affiliation{George P.\ and Cynthia Woods Mitchell Institute for Fundamental Physics and Astronomy, Texas A\&M University, College Station, TX, 77843-4242 USA}

\author[0000-0001-5414-5131]{Mark Dickinson}
\affiliation{NSF's National Optical-Infrared Astronomy Research Laboratory, 950 N. Cherry Ave., Tucson, AZ 85719, USA}

\author[0000-0001-5448-1821]{Robert Kennicutt}
\affiliation{Department of Physics and Astronomy, Texas A\&M University, College Station, TX, 77843-4242 USA}
\affiliation{George P.\ and Cynthia Woods Mitchell Institute for Fundamental Physics and Astronomy, Texas A\&M University, College Station, TX, 77843-4242 USA}
\affiliation{Department of Astronomy and Steward Observatory, University of Arizona, Tucson, AZ 85721, USA}

\author[0000-0002-8630-6435]{Anahita Alavi}
\affiliation{IPAC, California Institute of Technology, 1200 E. California Boulevard, Pasadena, CA 91125, USA}

\author[0000-0002-7959-8783]{Pablo Arrabal Haro}
\affiliation{NSF's National Optical-Infrared Astronomy Research Laboratory, 950 N. Cherry Ave., Tucson, AZ 85719, USA}

\author[0000-0002-9921-9218]{Micaela B. Bagley}
\affiliation{Department of Astronomy, The University of Texas at Austin, Austin, TX, USA}

\author[0000-0002-4193-2539]{Denis Burgarella}
\affiliation{Aix Marseille Univ, CNRS, CNES, LAM Marseille, France}

\author[0000-0002-9466-2763]{Aur{\'e}lien Le Bail}
\affiliation{Universit{\'e} Paris-Saclay, Universit{\'e} Paris Cit{\'e}, CEA, CNRS, AIM, 91191, Gif-sur-Yvette, France}

\author[0000-0002-5564-9873]{Eric F.\ Bell}
\affiliation{Department of Astronomy, University of Michigan, 1085 S. University Ave, Ann Arbor, MI 48109-1107, USA}

\author[0000-0001-7151-009X]{Nikko J. Cleri}
\affiliation{Department of Physics and Astronomy, Texas A\&M University, College Station, TX, 77843-4242 USA}
\affiliation{George P.\ and Cynthia Woods Mitchell Institute for Fundamental Physics and Astronomy, Texas A\&M University, College Station, TX, 77843-4242 USA}

\author[0000-0002-6348-1900]{Justin Cole}
\affiliation{Department of Physics and Astronomy, Texas A\&M University, College Station, TX, 77843-4242 USA}
\affiliation{George P.\ and Cynthia Woods Mitchell Institute for Fundamental Physics and Astronomy, Texas A\&M University, College Station, TX, 77843-4242 USA}

\author[0000-0001-6820-0015]{Luca Costantin}
\affiliation{Centro de Astrobiolog\'ia (CAB), CSIC-INTA, Ctra de Ajalvir km 4, Torrej\'on de Ardoz, 28850, Madrid, Spain}

\author[0000-0002-6219-5558]{Alexander de la Vega}
\affiliation{Department of Physics and Astronomy, University of California, 900 University Ave, Riverside, CA 92521, USA}

\author[0000-0002-3331-9590]{Emanuele Daddi}
\affiliation{Universit{\'e} Paris-Saclay, Universit{\'e} Paris Cit{\'e}, CEA, CNRS, AIM, 91191, Gif-sur-Yvette, France}

\author[0000-0002-7631-647X]{David Elbaz}
\affiliation{Universit{\'e} Paris-Saclay, Universit{\'e} Paris Cit{\'e}, CEA, CNRS, AIM, 91191, Gif-sur-Yvette, France}

\author[0000-0001-8519-1130]{Steven L. Finkelstein}
\affiliation{Department of Astronomy, The University of Texas at Austin, Austin, TX, USA}

\author[0000-0001-9440-8872]{Norman A. Grogin}
\affiliation{Space Telescope Science Institute, Baltimore, MD, USA}

\author[0000-0002-4884-6756]{Benne W. Holwerda}
\affil{Physics \& Astronomy Department, University of Louisville, 40292 KY, Louisville, USA}

\author[0000-0001-9187-3605]{Jeyhan S. Kartaltepe}
\affiliation{Laboratory for Multiwavelength Astrophysics, School of Physics and Astronomy, Rochester Institute of Technology, 84 Lomb Memorial Drive, Rochester, NY 14623, USA}

\author[0000-0002-5537-8110]{Allison Kirkpatrick}
\affiliation{Department of Physics and Astronomy, University of Kansas, Lawrence, KS 66045, USA}

\author[0000-0002-6610-2048]{Anton M. Koekemoer}
\affiliation{Space Telescope Science Institute, 3700 San Martin Dr., Baltimore, MD 21218, USA}

\author[0000-0003-1581-7825]{Ray A. Lucas}
\affiliation{Space Telescope Science Institute, 3700 San Martin Drive, Baltimore, MD 21218, USA}

\author[0000-0002-6777-6490]{Benjamin Magnelli}
\affiliation{Universit{\'e} Paris-Saclay, Universit{\'e} Paris Cit{\'e}, CEA, CNRS, AIM, 91191, Gif-sur-Yvette, France}

\author[0000-0001-5846-4404]{Bahram Mobasher}
\affiliation{Department of Physics and Astronomy, University of California, 900 University Ave, Riverside, CA 92521, USA}

\author[0000-0003-4528-5639]{Pablo G. P\'erez-Gonz\'alez}
\affiliation{Centro de Astrobiolog\'{\i}a (CAB), CSIC-INTA, Ctra. de Ajalvir km 4, Torrej\'on de Ardoz, E-28850, Madrid, Spain}

\author[0000-0002-0604-654X]{Laura Prichard}
\affiliation{Space Telescope Science Institute, Baltimore, MD, USA}

\author[0000-0002-9946-4731]{Marc Rafelski}
\affiliation{Space Telescope Science Institute, 3700 San Martin Dr., Baltimore, MD 21218, USA}
\affiliation{Department of Physics and Astronomy, Johns Hopkins University, Baltimore, MD 21218, USA}

\author[0000-0002-9415-2296]{Giulia Rodighiero}
\affiliation{Department of Physics and Astronomy, Università degli Studi di Padova, Vicolo dell’Osservatorio 3, I-35122, Padova, Italy}
\affiliation{INAF - Osservatorio Astronomico di Padova, Vicolo dell’Osservatorio 5, I-35122, Padova, Italy}

\author[0000-0003-3759-8707]{Ben Sunnquist}
\affiliation{Space Telescope Science Institute, 3700 San Martin Drive, Baltimore, MD 21218, USA}

\author[0000-0002-7064-5424]{Harry I. Teplitz}
\affiliation{IPAC, Mail Code 314-6, California Institute of Technology, 1200 E. California Blvd., Pasadena CA, 91125, USA}

\author[0000-0002-9373-3865]{Xin Wang}
\affiliation{School of Astronomy and Space Sciences, University of the Chinese Academy of Sciences (UCAS), Beijing 100049, China}

\author[0000-0001-8156-6281]{Rogier A. Windhorst} 
\affiliation{School of Earth and Space Exploration, Arizona State University,
Tempe, AZ 85287-1404, USA}

\author[0000-0003-3466-035X]{{L. Y. Aaron} {Yung}}
\altaffiliation{NASA Postdoctoral Fellow}
\affiliation{Astrophysics Science Division, NASA Goddard Space Flight Center, 8800 Greenbelt Rd, Greenbelt, MD 20771, USA}


\begin{abstract}
We test the relationship between UV-derived star formation rates (SFRs) and the 7.7 \(\mu m\) polycyclic aromatic hydrocarbon (PAH) luminosities from the integrated emission of galaxies at $z\sim 0 - 2$. We utilize multi-band photometry covering 0.2 -- 160 $\mu$m from HST, CFHT, JWST, Spitzer, and Herschel for galaxies in the Cosmic Evolution Early Release Science (CEERS) Survey. We perform spectral energy distribution (SED) modeling of these data to measure dust-corrected far-UV (FUV) luminosities, $L_\mathrm{FUV}$, and UV-derived SFRs. We then fit SED models to the JWST/MIRI 7.7 -- 21~$\mu$m CEERS data to derive rest-frame 7.7~$\mu$m luminosities, $L_{770}$, using the average flux density in the rest-frame MIRI F770W bandpass. We observe a correlation between $L_{770}$ and $L_\mathrm{FUV}$, where $\log L_{770} \propto (1.27\pm0.04) \log L_\mathrm{FUV}$.  $L_{770}$ diverges from this relation for galaxies at lower metallicities, lower dust obscuration, and for galaxies dominated by evolved stellar populations. We derive a ``single--wavelength'' SFR calibration for $L_{770}$  which has a scatter from model estimated SFRs ($\sigma_{\Delta\mathrm{SFR}}$) of 0.24 dex. We derive a ``multi--wavelength'' calibration for the linear-combination of the observed FUV luminosity (uncorrected for dust) and the rest-frame 7.7~$\mu$m luminosity, which has a scatter of $\sigma_{\Delta\mathrm{SFR}}$=0.21 dex. The relatively small decrease in $\sigma$ suggests this is near the systematic accuracy of the total SFRs using either calibration. These results demonstrate that the rest-frame 7.7~$\mu$m emission constrained by JWST/MIRI is a tracer of the SFR for distant galaxies to this accuracy, provided the galaxies are dominated by star-formation with moderate-to-high levels of attenuation and metallicity. 
\end{abstract}

\keywords{Star Formation (1569)  ---  Polycyclic aromatic hydrocarbons (1280) --- James Webb Space Telescope (2291) --- Hubble Space Telescope (761) --- Galaxy evolution (594)  --- Infrared galaxies(790)}

\section{Introduction} \label{sec:intro}
Measuring the rate that galaxies form stars (the star-formation rate, SFR) remains a challenge in astrophysics. SFRs are not measured directly, but rather estimated based on observations of the direct or reprocessed light produced by young stars. In turn this estimation is extrapolated to a total SFR based on assumptions of the stellar initial mass function 
\citep[IMF, see reviews by][]{Kennicutt_1998, Kennicutt_2012}. There have been a number of empirically derived SFR
calibrations making use of the continuum or emission lines that have 
demonstrated to be indicative of the short-lived stellar populations in galaxies \citep[e.g.,][]{Calzetti_2007,Houck_2007,Kennicutt_2009,Hern_n_Caballero_2009,Hao_2011,Shipley_2016,Xie_2019,Cleri_2022}. Such tracers of star formation range from the 
X-ray to the radio, and exhibit varying ability to estimate SFRs without large uncertainties (with calibration systematics on the order of 30\%, \citealt{Kennicutt_1998}), where at least some of this is 
contingent upon the properties of a galaxy \citep[e.g stellar mass, optical depth, etc., ][]{Kennicutt_2012}. Understanding the galaxy properties that limit the accuracy of SFR 
tracers is crucial. Frequently employed SFR tracers such as \(H\alpha\), Far-UV (FUV), 
Near-UV (NUV), and even the widely recognized Pa\(\alpha\) suffer from attenuation by dust  \citep[][]{Calzetti_1994, Papovich_2009}, 
which can reduce the certainty of SFR estimates when 
employed for dust obscured galaxies \citep[][]{Kennicutt_1998}. 
\par 
Another complication is that most of the stellar light from galaxies at cosmic noon is absorbed and then emitted again at longer wavelengths, where obscured galaxies contribute up to \(\sim\)80\% of the star-forming population for \(z \sim1-
3\) \citep{Madau_2014}. As such, calibrations for tracers 
that are capable of measuring SFRs for obscured galaxies 
are essential for discerning the overall picture of star 
formation during these epochs. The infrared (IR) has been frequently 
employed for this purpose in the obscured population of galaxies, more specifically 
the total IR luminosity ($L_{\mathrm{TIR}}$= 3\(\mu m\) -- 1100\(\mu m\)), the mid-IR polycyclic 
aromatic hydrocarbons (PAHs), and the 24 \(\mu m\) feature \citep[see, e.g.,][]{Kennicutt_2012}. Tracers that rely on far-IR data have not had much recent studies since the end of far-
IR space based missions such as Spitzer, Herschel, 
the Infrared Astronomical Satellite (IRAS), and the Infrared Space 
Observatory (ISO). As JWST \citep[][]{Gardner_2006} continues to unveil new dust obscured galaxies, there is a growing need for new methods to study star formation. \par 

The mid-IR offers new opportunities to explore star formation in dust obscured galaxies, which is imperative for understanding the new era of galaxies uncovered with JWST. The mid-IR covers strong PAH emission features at rest-frame wavelengths 3 -- 18 \( \mu m\), and are found in photo-dissociation regions surrounding HII regions \citep[][]{Calzetti_2007, Smith_2007}. PAHs contribute up to 20\% of the total-IR luminosity for star-forming galaxies \citep[][]{Elbaz_2011}, with the 7.7 \(\mu m\) feature consisting of up to half of the total PAH luminosity \citep[][]{Smith_2007}. The 7.7 \(\mu m\) PAH emission has been shown in previous works to correlate with the SFR for resolved star-forming regions \citep[][]{Calzetti_2007} and for the integrated emission of galaxies  \citep[]{Houck_2007, Pope_2008, Hern_n_Caballero_2009, Pope_2013, Cluver_2014, Shipley_2016, Xie_2019}. However, the correlation between the PAH  emission and the SFR at z \(>\) 1 for large samples remain an unexplored territory in literature due to the sensitivity of available IR instruments prior to the launch of JWST. \par

The JWST mid-IR instrument  \citep[MIRI,][]{Wright_2023} is sensitive to the emission from galaxies at wavelengths $\sim 5-28$~\(\mu m\), including those from the PAH features at an unprecedented level. Results from the first year of JWST highlight the capability of MIRI to trace the PAH features and constrain the 7.7 \(\mu m\) PAH emission \citep[e.g.,][]{Chastenet_2022, Evans_2022, Dale_2023, Kirkpatrick_2023,Shen_2023, Yang2023}. The 7.7 \(\mu m\) emission can be observed up to a redshift of 2 with the available MIRI bands, and can achieve depths up to two orders of magnitude fainter than Spitzer. There have yet to be galaxy-scale star formation (SF) studies done for the 7.7 \(\mu m\) with MIRI, with the exception of preliminary tests from  \cite{Shipley_2016} and \cite{Kirkpatrick_2023}. MIRI observations allow us to probe the faint end of the relation between the 7.7 $\mu m$ feature and SFR that has yet to be observed by any of the JWST predecessors out to z $\sim$ 2.\par 
This paper presents one of the first tests of the rest-frame 7.7 \(\mu m\) PAH emission using JWST/MIRI imaging to track the SFR on galaxy--wide scales for sources at redshifts 0 -- 2. To do this, we perform spectral energy distribution (SED) modeling with multi-band photometry from the UV to far-IR in addition to MIRI photometry from the Cosmic Evolution Early Release Science (CEERS) survey \citep{Finkelstein_2017, Yang2023}. We use the model SEDs to measure the rest-frame observed FUV luminosity (un-corrected for dust), FUV attenuation ($A(\mathrm{FUV})$), stellar mass, and SFR. In addition, we perform SED modeling of the CEERS JWST/MIRI 7.7 -- 21 $\mu m$ data to measure the rest-frame 7.7 \(\mu m\) luminosity measured using the average flux-density in the MIRI F770W bandpass (\(L_{770}\)). We compare the correlation between the dust-corrected FUV luminosity and \(L_{770}\), and use the relation to derive a ``single--wavelength'' SFR calibration. We then model the dust corrected FUV luminosity using a linear combination of the observed FUV luminosity and \(L_{770}\) for a ``multi--wavelength'' SFR calibration.
\par

The outline of this paper is as follows. Section \ref{sec:data} 
provides an overview of our UV, optical, mid-IR, and far-IR imaging 
and multi--wavelength catalogs. We describe our selection methods 
and sample in Section \ref{sec:sample}. Section \ref{sec:Methods} 
describes our SED modeling and measurements of the  rest-
frame \(L_{770}\) and observed FUV luminosities, as well as our 
prescription for attenuation. In Section \ref{sec:results} we show 
our results and discuss their implications in addition to caveats in 
Section \ref{sec:discussion}. Lastly, our summary and main conclusions are in Section \ref{sec:summary_concl}. Throughout this
paper all magnitudes are presented in the AB system \citep[][]{Oke1983, Fukugita1996}. We use the standard Lambda cold dark matter (\(\Lambda\)CDM) cosmology with \(H_0\) = 70 km \({Mpc}^{-1} s^{-1}\), \(\Omega_\Lambda\) = 0.70, and \(\Omega_M\) = 0.30. \par

\section{Data} \label{sec:data}
\subsection{The Cosmic Evolution Early Release Science (CEERS) Survey}
CEERS \citep[][]{Finkelstein_2017} (Proposal ID \#1345) 
covers $\sim$100 sq.\ arcmin with JWST imaging and 
spectroscopy, targeting the Extended Groth 
Strip  \citep[EGS,][]{Davis_2007}. The June 2022 
observations include four MIRI and NIRCam 
pointings taken in parallel. Pointings CEERS MIRI 1 and 
CEERS MIRI 2 used for this work have no overlap with the CEERS NIRCam imaging, and were covered with 
MIRI in six continguous filters:  F770W, F1000W, F1280W, F1500W, 
F1800W and F2100W (see \citealt{Bagley_2023} and \citealt{Yang2023}). Importantly, these two fields have overlap with the Cosmic Assembly Near-
infrared Deep Extragalactic Legacy Survey \citep[CANDELS,][]{Grogin_2011, Koekemoer_2011} and they are the only two such CEERS MIRI fields that have follow-up UV imaging from Hubble Space Telescope (HST) as part of the UVCANDELS program \citep[]{Wang_2020}. This provides coverage from 0.2 -- 1.8~\micron\ with HST-quality angular resolution, which is well matched to MIRI (see below).  The other two CEERS MIRI pointings (3 and 6) observed deeply with the F560W and F770W to study the stellar populations of more distant galaxies ($z > 4$, \citealt{papovich2023ceers}) and only trace the 7.7 \(\mu m\) PAH emission for very low redshift galaxies.
\par

\subsubsection{CEERS MIRI Imaging and Photometry Catalog}\label{subsec:jwstmiriobs}
A full description of the MIRI data and reduction is provided in \cite{Yang2023}, but will briefly be described here \citep[also see][]{Yang_2021, Kirkpatrick_2023,Shen_2023, Yang_2023}.\par

\begin{figure*}[ht]
\centering
\includegraphics[width=.65\textwidth]{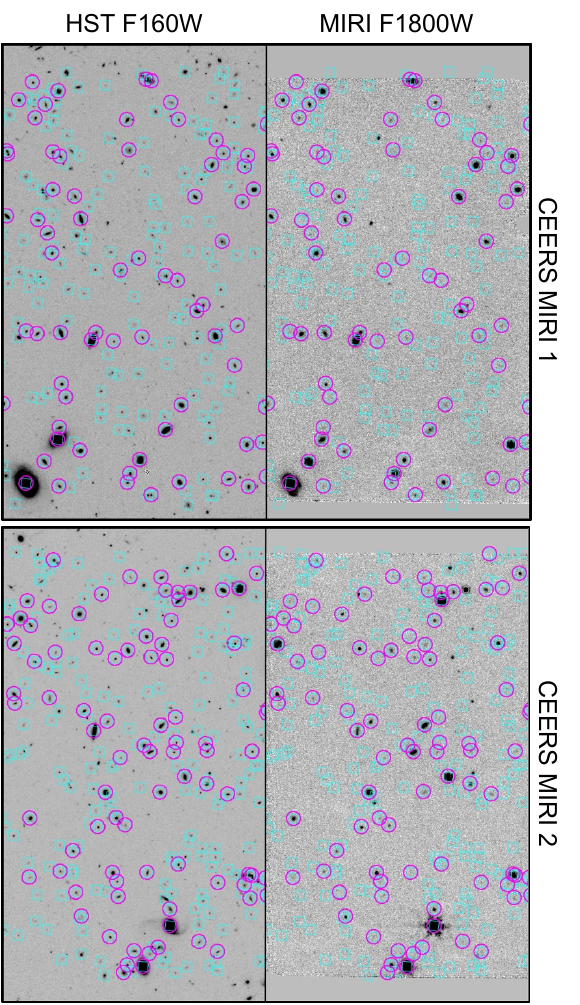}
\caption{HST F160W (left) and MIRI F1800W (right) for fields CEERS MIRI 1 (top row) and CEERS MIRI 2 (bottom row). Blue squares are sources with m(F160W) \( < \) 26.6 mag at z $< 2$, and purple circles (diameter of 2.6" for scale) are sources with S/N(rest-frame \(7.7 \mu m\)) \(> 4\). \label{fig:HSTvsMIRI}}
\end{figure*}

\begin{figure*}[t]
\centering
\includegraphics[scale=0.55]{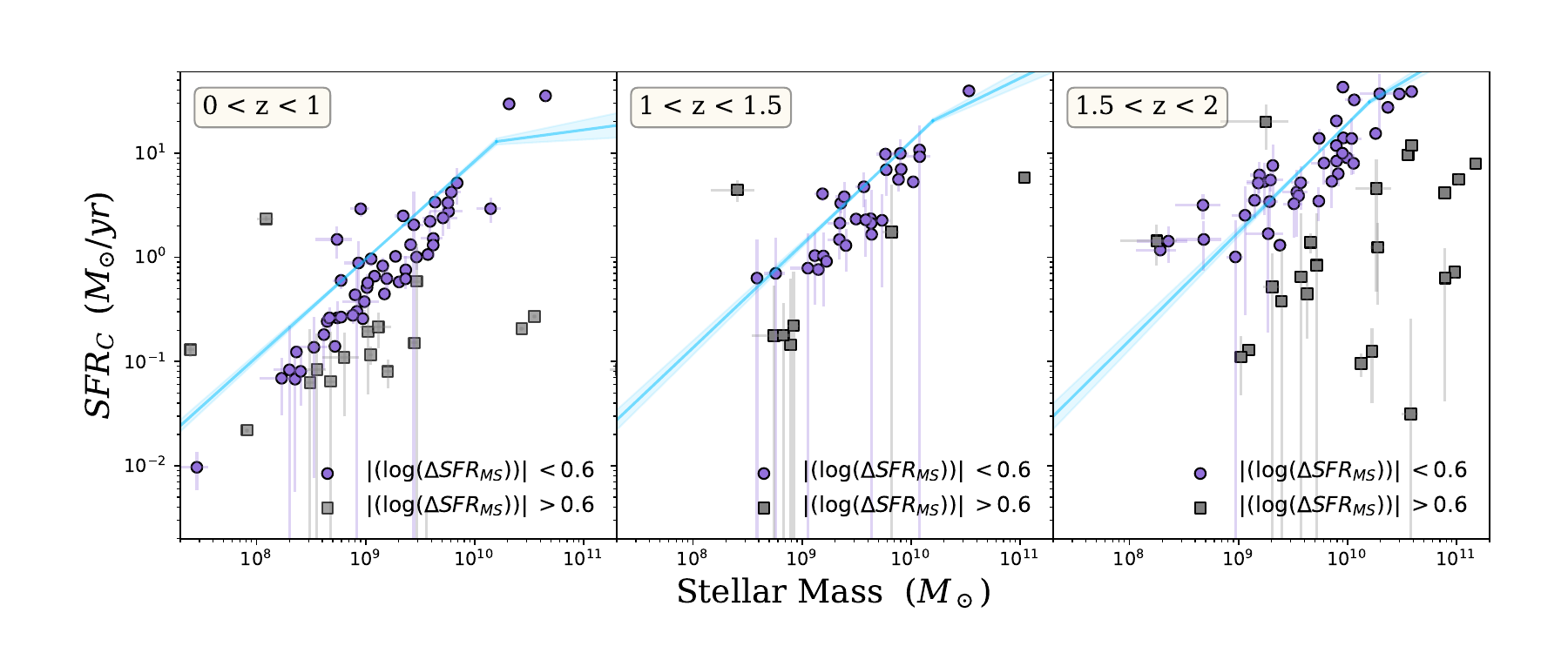}
\caption{SFR--mass relation for the galaxies in our samples compared to the star-forming main sequence. The panels show the SED-measured SFRs from \textsc{CIGALE} ($SFR_C$, section \ref{subsec:sedgen}) as a function of the estimated stellar masses in different redshift bins (as labeled). The blue line indicates the star--forming main sequence and the uncertainty in the relation in the blue shaded region from \cite{Whitaker2014}.}\label{fig:SFMS}
\end{figure*}

The MIRI images were processed with the \textsc{JWST Calibration Pipeline} (v1.10.2) \citep{bushouse_howard_2022_7038885} primarily using the default parameters for stages 1 and 2. 
The background was modeled with the median of all images in the same bandpass but different fields and/or dither 
positions. With the background subtracted from each image, the astrometry is corrected by matching to CANDELS imaging 
\citep[as described in more detail in][]{Bagley_2023} prior to stage 3 processing in the pipeline. This results in the final science images, weight maps, and uncertainty images with a pixel scale of 0.09 "/pix  registered to the CANDELS v1.9 WFC3 images. This last step is important as we use MIRI fluxes matched to the existing CANDELS HST catalogs. 

The MIRI photometry is measured from sources selected from the CANDELS WFC3 catalog \citep[][and see below]{Stefanon_2017} with T-PHOT \citep[v2.0, ][]{Merlin_2016}. T-PHOT uses priors from the HST/WFC3 F160W for the lower resolution MIRI images for the photometric analysis. The point spread function (PSF) for each MIRI band is constructed using WebbPSF \citep{Perrin2012}. The kernels are then constructed to match the PSF from the CANDELS/WFC3 F160W image (FWHM of \(\simeq\) 0.2")) to the MIRI images (FWHM of \(\simeq\) 0.2" - 0.5")) to extract source photometry with T-PHOT. These fluxes and their uncertainties for each source are used as the MIRI catalog. We note that the MIRI flux calibration has since been updated from \cite{Yang2023} where the median offset in MIRI F770W is 0.18 mag and is substantially lower for the redder MIRI bands, which are impacted by only a few percent \footnote{Please visit the following \href{https://www.stsci.edu/contents/news/jwst/2023/temporal-behavior-of-the-miri-reduced-count-rate.html}{link for more.}}.
\par 

\subsection{CANDELS Imaging and multi--wavelength Photometry Catalog}\label{subsec:CANDELdata}
We use the catalog from \cite{Stefanon_2017}, which provides matched-aperture photometry for a larger range of observations covering 0.4 -- 8 \(\mu m\) in the EGS field built upon the CANDELS, All-wavelength 
Extended Groth strip International Survey (AEGIS), and 3D-HST 
program with imaging from Canada France Hawaii Telescope 
(\emph{CFHT})/MegaCam, NEWFIRM/NEWFIRM, \emph{CFHT}/WIRCAM, HST/ACS,
HST/WFC3, and Spitzer/IRAC.  The catalog also includes photometric redshifts and estimated properties from SED fitting of the multi--wavelength photometry, which were 
independently carried out by 10 different groups, each using 
different codes and/or SED templates including FAST 
\citep[][]{Kriek2018}, HyperZ \citep[][]{Bolzonella2000}, Le Phare 
\citep[][]{Ilbert2006}, WikZ \citep[][]{Wiklind2008}, SpeedyMC 
\citep[][]{Acquaviva2012}, and other available codes 
\citep[][]{Fontana2000, Lee2010}. This catalog includes the 
spectroscopic and photometric redshifts that will be used for 
this work. The spectroscopic redshifts in this catalog are from 
the DEEP2/DEEP3 surveys \citep[][]{Coil_2004,Willner_2006,Cooper_2011,Cooper2012,Newman_2013}, and the photometric 
redshifts are measured using the methods outlined from \cite{Dahlen_2013} and 
\cite{Mobasher_2015}.\par 

\subsection{UV-CANDELS Imaging} \label{subsec:UVCANDELdata}
We also use the HST WFC/F275W and ACS/F435W imaging as part of the Ultraviolet Imaging of a portion of the CANDELS field from UVCANDELS, a Hubble Treasury program (GO-15647, PI: H. Teplitz). The primary UVCANDELS WFC3/F275W imaging reached m \(\leq\) 27 mag for compact galaxies (corresponding to a SFR \(\sim\) 0.2 \(M_\odot yr^{-1}\)) at $z = 1$, and the coordinated parallel ACS/F435W imaging reached m \(\leq\) 28 mag. A UV optimized aperture photometry based on optical isophotes aperture was utilized, similar to the work done on the Hubble Ultra-Deep Field UV analysis \citep{Teplitz_2013, Rafelski_2015}. The smaller optical apertures without degradation to the image quality allows the UV-optimized photometry method to reach the expected \(5 \sigma\) point-source depth of 27 mag in F275W. The UV-optimized photometry yields a factor of \(\sim 1.5 \times\) increase in signal-to-noise ratio in the F275W band with higher increase in brighter extended objects, which complement the pre-existing CANDELS multi--wavelength catalog from \citep[][]{Stefanon_2017}.\par 

\subsection{Spitzer and Herschel Far-IR Data Imaging}
We also use Spitzer MIPS 24 \(\mu m\) and Herschel PACS 100 and 160 \(\mu m\) photometry for the EGS field from the ``super-deblended'' catalog of Le Bail et al.\citep[in prep, as briefly described in][]{LeBail_2023}. This catalog was developed following the methods outlined in \cite{Jin2018} and \cite{Liu2018} for COSMOS and GOODS-N respectively, where specifically the MIPS and PACS photometry were extracted by PSF fitting to prior positions of galaxies from the \cite{Stefanon_2017} catalog.  \par 

\section{Sample\label{sec:sample}}
\subsection{Selection Criteria \label{subsec:sample_sel}}
We use as our initial sample all coordinate matched galaxies in 
the UVCANDELS and MIRI catalogs in fields CEERS MIRI 1 and CEERS MIRI 2 with 
F160W magnitude \(<\) 26.6 mag (i.e., at the 90\% completeness 
limit, see \citealt{Shen_2023}). We select these fields as they 
are the only two pointings that overlap with the UVCANDELS data 
with the full complement of MIRI bands from 7.7 to 21~$\mu$m. We 
further restrict our sample to have $z < 2$ in order to ensure 
that the 7.7 $\mu$m PAH feature falls within the wavelength 
coverage of the MIRI bands. This work utilizes spectroscopic 
redshifts when available from \cite{Stefanon_2017} and compiled from the literature (N. Hathi, private communication). After visually inspecting the faintest objects 
in the MIRI mosaics, we determine that robust detections are possible for objects with S/N $>$ 4.  We then require that objects have detections more significant than 4\(\sigma\)  in the MIRI 
band in which the 7.7 \(\mu m\) PAH feature resides (i.e., the band that includes $(1+z) \times 7.7$ $\mu m$). On average with MIRI we detect one-third of the sources detected by HST/F160W in the redshift range $0 < z < 2$ (see Figure~\ref{fig:HSTvsMIRI}). In addition to identifying faint objects in the MIRI mosaics, we also flag galaxies 
that either reside on the edge of the image (where there is less exposure 
time) or that experience contamination from bright sources or diffraction spikes. Such galaxies are marked with 
\texttt{mosaicFlag} = 1.  We then only select galaxies with 
\texttt{mosaicFlag} = 0 for this work. We also identify and 
remove galaxies with AGN in our sample by using the IRAC 
color selection from \cite{Donley_2012}.  \par

Lastly, we apply a selection to ensure we include only actively star-forming galaxies that are along the star-forming main 
sequence,  as defined by \cite{Whitaker2014}. Using the SFR and stellar mass estimates from our modeling of the SEDs (see Section \ref{subsec:sedgen} below), we measure $\Delta SFR_{MS}$ defined as $| \log(SFR_{MS}) - \log(SFR_C)|$. $SFR_C$ is the model--estimated SFR from the SED fitting with \textsc{CIGALE} as described in Section \ref{subsec:sedgen} (the subscript ``C'' stands for \textsc{CIGALE}).  $SFR_{MS}$ is the value of the main-sequence SFR from \cite{Whitaker2014} at a fixed stellar mass and redshift. Through visual inspection of our sources along the star-forming main sequence (Figure \ref{fig:SFMS}), we determine that $\Delta SFR_{MS} < 0.6$~dex is sufficient to select only star-forming galaxies in our sample (that is, we select galaxies that have a SFR within a factor of $\approx$4 of the SFR$_{MS}$).  This selection removes both quenching/quenched galaxies (those below SFR$_{MS}$) and galaxies in a ``burst'' that lie above SFR$_{MS}$. Starbursts can also lie within the star-forming main sequence \citep[see][]{Elbaz_2018, Gomez_Guijarro_2022}, which can be diagnosed using the ratio of the total IR luminosity and the PAH luminosity. We cannot consider such sources here as less than 1\% of our sources have coverage in the far-IR (at $\sim$70--160 $\mu m$), and we defer the analysis of these objects for a future study. A summary of all our selection criteria as well as the number of galaxies in our sample are listed in Table \ref{tab:sample_selection}.\par

\begin{deluxetable}{cc}[ht]
\tablecaption{Summary of Sample Selection \label{tab:sample_selection}}
\tablewidth{0pt}
\tablehead{
\colhead{Selection Criteria} & \colhead{\# of Galaxies}
}
\startdata
 Initial Sample with m(F160W) \( < \) 26.6 mag  & 816 \\[0.13cm]
\hline 
Redshift range: \(0 < z < 2\)  & 607 \\[0.13cm]
\hline 
MIRI brightness: S/N(rest-frame \(7.7 \mu m\)) \(> 4\) & 189 \\[0.13cm]
\hline 
Sufficient coverage: MIRI \texttt{mosaicFlag} = 0  & 173 \\[0.13cm]
\hline 
Not an AGN: AGN flag = 0 & 173 \\ [0.13cm]
\hline 
Final sample with |(log($\Delta SFR_{MS}$))| \( < 0.6\) & 120 \\ [0.13cm]
 \hline \hline
\enddata
\end{deluxetable}

\subsection{Sample Properties \label{subsec:sample_desc}}
\begin{figure*}[ht]
\centering
\includegraphics[scale=0.5]{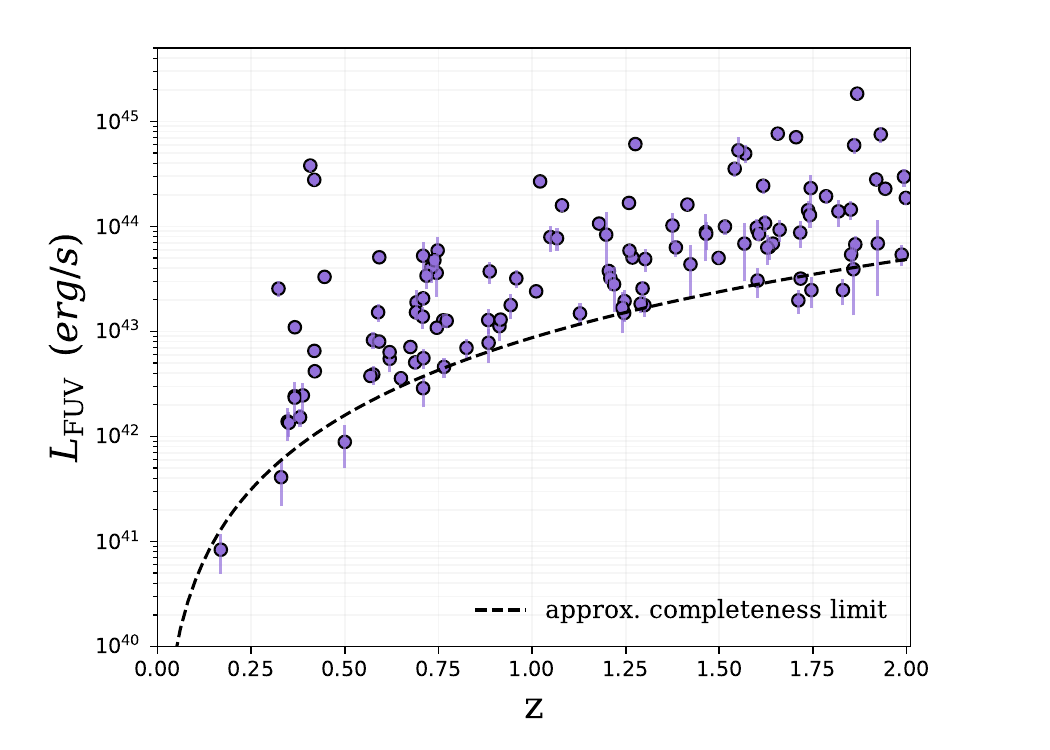}
\includegraphics[scale=0.5]{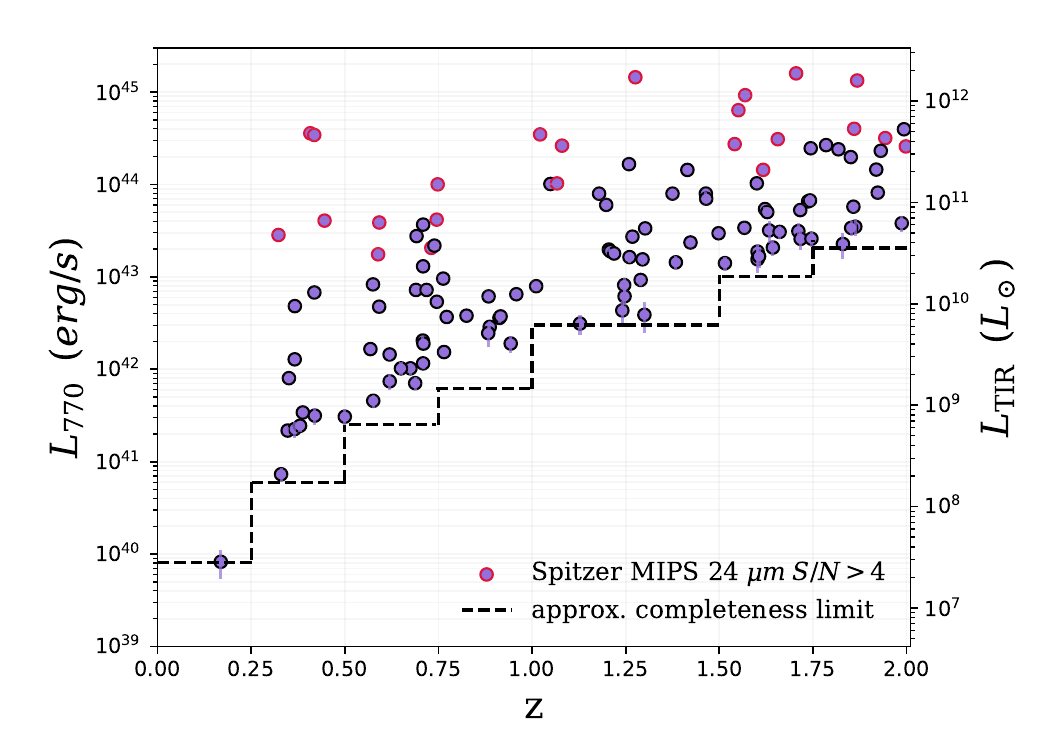}
\caption{(Left:) Dust-corrected UV luminosity $L_{\mathrm{FUV}}$ compared to redshift for our final sample. The black dashed line shows the approximate 90\% completeness limit of the UV-CANDELS data. (Right:) \(L_{770}\) versus redshift for our final sample. The right ordinate shows the corresponding $L_{\mathrm{TIR}}$ estimated from our SED modeling. The black dashed line in each redshift bin indicates the \(4\sigma\) limit with MIRI. \label{fig:samplecompleteness}}
\end{figure*}

Figure 
\ref{fig:samplecompleteness} shows the distribution of 
the rest-frame 7.7 $\mu m$ luminosity ($L_{770}$) and dust-corrected FUV luminosity ($L_\mathrm{FUV}$, see 
Section \ref{subsec:sedgen} and \ref{subsec:sedint}) as a function of redshift for the 120 galaxies in our final sample (see Table~\ref{tab:sample_selection}). The majority (87\%) of galaxies in our sample reside at $z > 0.5$, and the majority (89\%) have inferred $L_\mathrm{TIR}$ values above $10^9$~$L_\odot$. The final sample includes a large range of dust--attenuation in the visual ($A_V$, estimated from the \textsc{CIGALE} SED fits in Section \ref{subsec:sedgen}), ranging from $A_V$ = 0.35 mag to 3 mag.  The MIRI data detect the mid-IR emission of galaxies at much fainter flux densities than previous instruments.  For example, only four out of 15 galaxies in our sample at z $ > 1.8$ have a Spitzer/MIPS 24 $
\mu m$ detections with S/N $>4$. This is relevant because it is at this redshift where the 7.7 $\mu m$ feature would have been observed by Spitzer at 24 $\mu m$. These four galaxies have an average $L_\mathrm{TIR}$ of roughly $7 \times 10^{11}$~$L_\odot$, whereas at the same redshifts MIRI is sensitive to the mid-IR emission from sources down to $ 4 \times 10^{10}$~$L_\odot$, more than an order of magnitude fainter. With increased sensitivity from MIRI, we are able to observe SFR$_C$ as faint as $10^{-2} \;  M_\odot yr^{-1}$ up to  $\sim 10^{2} \;  M_\odot yr^{-1}$. With this, our final sample encompasses a wide range of galaxies with considerably varying star-forming activity, dust obscuration, and total IR luminosities for redshifts up to 2. We explore 
how different properties of our sample (e.g., mass or attenuation) impact the ability of PAH to trace SF in Section \ref{sec:discussion}. \par

\section{Methods} \label{sec:Methods}

\subsection{SED Modeling}\label{subsec:sedgen}
We model the SEDs built from the multi--wavelength photometry for our sample with \textsc{CIGALE} \citep[][]{Boquien_2019}. \textsc{CIGALE} uses simple stellar populations and parametric star formation histories to build composite stellar populations. The code then calculates the emission from ionized gas and thermal dust emission that is balanced based on the dust attenuation from flexible attenuation curves. The \texttt{dl2014} module used for this work considers a multi-component dust emission based on the \cite{Draine2014} models. The first dust emission component considers heating from a single source such as a stellar population, whereas the second considers variable heating linked to star-forming regions \citep{Boquien_2019}. \textsc{CIGALE} uses a large grid of models that is fitted to the data where the physical properties are then estimated by analyzing the likelihood distribution. \par 

\begin{figure*}[t]
\centering
\includegraphics[width=.6\textwidth]{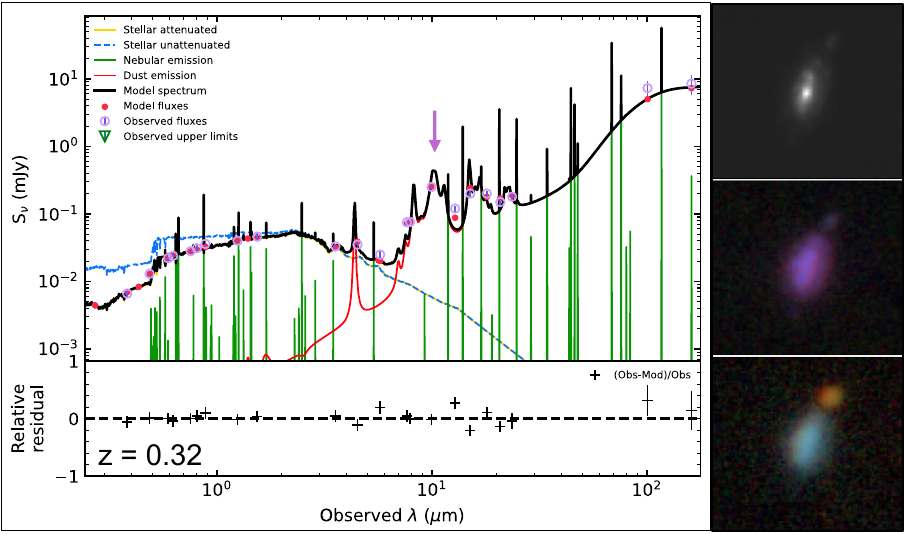}
\includegraphics[width=.6\textwidth]{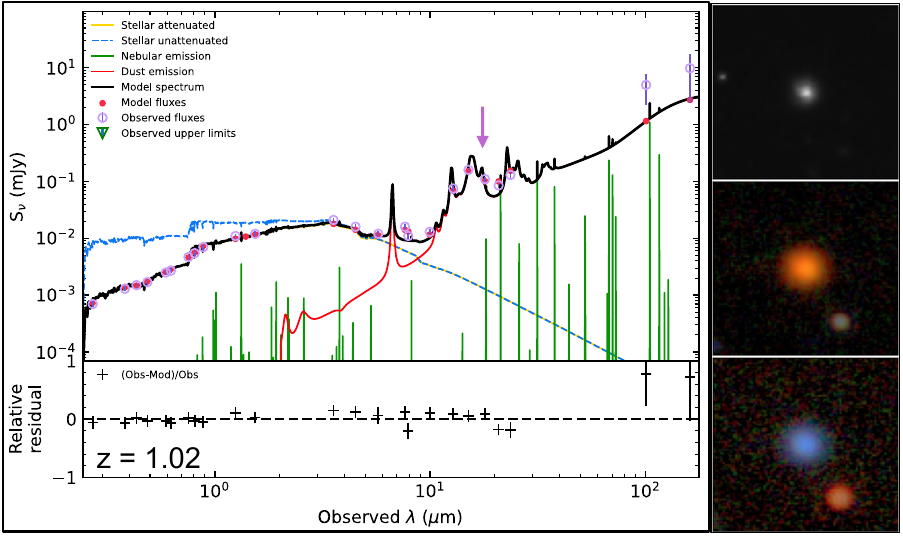}
\includegraphics[width=.6\textwidth]{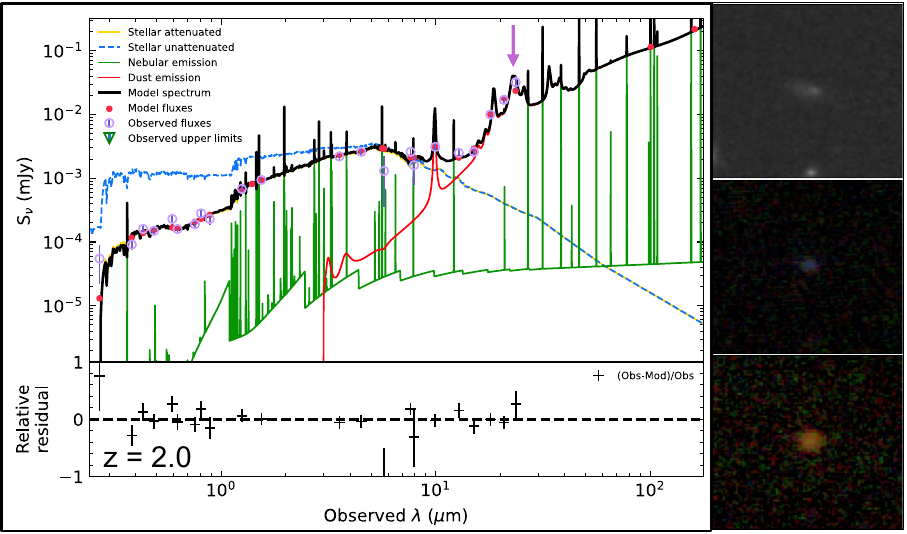}
\caption{Each panel shows an example of the best-fit SEDs from \textsc{CIGALE} for sources in 
the selected sample in order of increasing redshift (top to bottom).  The data in each plot show the measured flux densities with each curves representing different model components (see panel legend). The location of the observed 7.7 $\mu m$ feature is highlighted in each panel by the purple arrow. To the right of each panel are \( 10 \arcsec \times 10 \arcsec\)  
postage stamps of the CEERS HST F160W (top), composite RGB image of JWST/MIRI 
F1000W (B) + F1280W (G) + F1500W (R)(middle), composite RGB image of JWST/MIRI 
F1500W (B) + F1800W (G) + F2100W (R) (bottom). The MIRI images shift in 
color as the 7.7~$\mu$m PAH feature redshifts through the different MIRI bands. At $z=0.32$, the  7.7~$\mu$m PAH feature falls in the F1000W band, with an additional PAH feature that peaks at F1500W which makes the image ''purple''.  At $z=1.02$, the 7.7~$\mu$m PAH feature falls in the F1500W band, making the middle image ``red'' 
and the bottom image ``blue''. At $z=2.0$, the PAH feature falls in the F2100W band, making the 
bottom image ``red''. \label{fig:MIRI_SEDs}}
\end{figure*}

We consider two different SED modeling cases for this study where we fit the data with \textsc{CIGALE}. For both cases we use the same parameters as shown in Table \ref{tab:CIGALEparam} at fixed redshifts. In 
Case 1, we fit models to the MIRI data \textit{only} for each object to test 
the capability of \textsc{CIGALE} to derive the rest-frame 7.7 \(\mu m\) luminosity. This is to ensure that the measured rest-frame 7.7 $\mu$m luminosity is not influenced by the rest of the galaxy multi--wavelength SED.\par  

In Case 2 we use all available photometry from UVCANDELS through the far-IR, which includes the MIRI data.  While we include far-IR data when available, only 23 of the 120 galaxies in our sample have a S/N $> 4$ detection from MIPS 24 $\mu m$, and only one source is detected by Hershel/PACS at 160 $\mu m$. Given that \textsc{CIGALE} is dependent on the principle of an energy balance, the MIRI data play a significant role for the Case 2 SED fitting.  We will visit the significance of the MIRI data on the Case 2 models further in Section \ref{subsec:caveats}. \par

We use the results from the Case 2 fits to measure the (1) observed FUV luminosity, defined by equation \ref{eqn:continuum_int},  (2) the  stellar mass, (3) the $SFR_C$, and (4) the estimated FUV dust attenuation ($A(\mathrm{FUV})$).  We note that \({SFR}_C\) is the SFR averaged over the previous 10 Myr using the fitted star-formation histories (we discuss this further in Appendix \ref{sec:modelestimates}).\par 

Figure \ref{fig:MIRI_SEDs} shows example \textsc{CIGALE} SED fits from Case 2 for several sources in our sample in order of increasing redshift.  For Case 1 and Case 2 models we measure the typical mean reduced $\chi^2$ for our final sample to be 1.65 and 1.53 respectively. \par
\par

\begin{deluxetable}{lll}[ht]
\tabletypesize{\footnotesize}
\tablecaption{\textsc{CIGALE} parameters used for SED fitting \label{tab:CIGALEparam}}
\tablewidth{0pt}
\tablehead{
\colhead{\raggedleft{Module}} & \colhead{Parameter} & \colhead{Input Value(s)}
}
\startdata
SFH: & \(\tau\) [Myr] & {0.1, 0.5, 1.0, 5.0}\\
SFR(t) \(\propto \frac{t}{\tau ^2} \times exp (\frac{t}{- \tau})\) & t [Myr] & {0.5, 1.0, 3.0, 5.0, 7.0} \\ \hline
Simple stellar population: & IMF & \cite{Chabrier2003} \\  
\cite{Bruzual2003}& Metallicity & $Z_\odot$ = 0.02  \\ \hline
Dust attenuation: & \(E(B-V)_l\) & {0.1, 0.3, 0.5, 0.8} \\ 
\cite{Calzetti2000} & UV bump amplitude & {0.0, 1.5, 3.0} \\ 
{    } & power-law slope & {$-$0.3, $-$0.1, 0.0} \\ \hline
Dust Emission: & \(q_{pah}\) & {0.47, 1.77, 3.19, 4.58,
} \\
\cite{Draine2014} & {} & { 5.95, 7.32}\\
{} &\(u_{min}\) & {0.2, 1.0, 5, 10, 20, 35} \\
{       }& \(\gamma\) & {0, 0.005, 0.1, 0.02} \\ \hline
Redshift & z & Fixed at z of  \\ 
{     } & {      } & \cite{Stefanon_2017} \\
{     } & {      } & and (N. Hathi, private \\
{     } & {      } &  communication) \\
\hline \hline
\enddata
\tablecomments{The default \textsc{CIGALE} values were used for parameters not listed in this table.}
\end{deluxetable}

\subsection{SED Integration and Luminosity Calculation}\label{subsec:sedint}
We calculate the average rest-frame flux in the MIRI F770W band (\(\langle F_{770} \rangle\)) as an approximation for the 7.7 \(\mu m\) brightness. The distinction in notation primarily serves as a reminder of our methodology for measuring the 7.7 \(\mu m\) flux in this study. To calculate \(\langle F_{770} \rangle\) we use,
\begin{equation} \label{eqn:integration}
\langle F_{770} \rangle = \frac{\int_{c/8.8\mu m}^{c/6.4\mu m} (T(\nu)F_{\nu})/{\nu}\ \,d{\nu} }{\int_{c/8.8\mu m}^{c/6.4\mu m} T(\nu)/{\nu}\ \,d{\nu}}.
\end{equation}
Where \(F_{\nu}\) and \(\nu\) are the flux and frequency from the model SEDs output from Case 1. \(T(\nu)\) is the MIRI F770W transmission filter taken from the Spanish Virtual Observatory \citep[][]{Rodrigo_2012, Rodrigo_2020}. We then measure the PAH luminosity as,
\begin{equation} \label{eqn:luminosity}
L_{770} =  4\pi{(D_L)^2}{\langle F_{770} \rangle}{{\nu}_{(7.7 \mu m)}}.
\end{equation}
Where $D_L$ is the luminosity 
distance, calculated with the \textit{astropy.cosmology} package in python, and $\nu_{7.7}$ is the rest-frame frequency at 7.7~$\mu$m ($\nu_{7.7} = c / 7.7~\mu$m). \par 

We note that the rest-frame 7.7 $\mu$m contains the bright emission from the 7.7 PAH complex \textit{and} the dust continuum (and we make no correction for the latter). \(L_{770}\) primarily does not have any contribution from the stellar continuum for star-forming galaxies, such as those in our sample. However, there are a few exceptions which we discuss in Section \ref{subsec:caveats}. The dust continuum is estimated to contribute up to 10\% of the emission compared to the 7.7 $\mu$m PAH luminosity for MIRI F770W \citep[][]{Chastenet_2022}. This value is an approximation for star-forming galaxies at low redshift from the SINGS sample, which will require further study with JWST/MIRI to determine if this holds for fainter galaxies at higher redshifts. With this, we will assume that \(\langle F_{770} \rangle\)  is dominated by the equivalent width of the 7.7~$\mu$m PAH emission in star-forming galaxies at z $<$ 2. We explore the impact of the dust continuum on our results in Section \ref{subsec:caveats} and Appendix \ref{sec:l770vsl77}.\par

We estimate the observed FUV luminosity following the same definition as \cite{Kennicutt_2012} with a central wavelength of 0.155 \(\mu m\) and \(\Delta \lambda = 0.2 \mu m\). We integrate the fitted models from \textsc{CIGALE} using, 
\begin{equation}\label{eqn:continuum_int}
    F_{\mathrm{FUV}_{obs}} = \int_{0.135 \mu m}^{0.175 \mu m} F_{\nu}\,d{\nu}.
\end{equation}
The observed FUV luminosity is then corrected for dust using the Case 2 model output FUV attenuation ($A(\mathrm{FUV})$). This yields,
\begin{equation}\label{eqn:fuv_luminosity}
    F_{\mathrm{FUV}} = F_{\mathrm{FUV}_{obs}} \times {10^{0.4 \times A(\mathrm{FUV})}}.
\end{equation}
Where the dust-corrected FUV luminosity is \(L_\mathrm{FUV} = 4\pi{(D_L)^2} F_{\mathrm{FUV}}\). \par 

\subsection{Estimate of Uncertainties on Derived Quantities}\label{subsec:error_estimation}
We estimate the uncertainties on derived quantities, including the rest-frame observed FUV and 7.7~$\mu$m luminosities using a Monte Carlo (MC) simulation. To do this, we perturb the photometry for each galaxy in the catalogs 1000 times with a random value ($R$) taken from a normal distribution with mean, $\mu=0$ and variance, $\sigma^2=1$, i.e., $N(\mu=0, \sigma^2=1)$, then multiplied by the observed errors, $f_\mathrm{err}$, and added to the measured flux density, $f$. This yields the following equation,
\begin{equation} \label{eqn:flux_pert}
    f_{new} = f + f_{err} \times R.
\end{equation}
In each 
iteration, we re-fit the perturbed galaxy SED using 
the same method outlined in Section 
\ref{subsec:sedgen}. Following the same equations in 
Section \ref{subsec:sedint}, we then measure \(L_{770}\) and the observed FUV luminosity from each newly modelled 
SED for each source. After the 1000 iterations are 
complete, we measure the 1\(\sigma\) standard 
deviation, which is 
used as our uncertainty estimate on these quantities. For uncertainty in the dust-corrected FUV luminosity (L$_{\mathrm{FUV}}$), we also consider the model estimated uncertainties in $A(\mathrm{FUV})$ and propagate accordingly for each source. \par

\section{Results} \label{sec:results}
In this Section we examine the correlation between the rest-frame dust-corrected FUV luminosity and the rest-frame 7.7 \(\mu m\) luminosity. Our goals for this section are to asses the application of the 7.7 $\mu$m luminosity as a tracer of star formation for the following cases: (1) retrieving information on the total SFR when only MIRI data is available and (2) the PAH luminosity as a tracer of obscured star formation when FUV data is also available. From the analysis we measure a ``single--wavelength'' calibration between the 7.7~$\mu$m luminosity and the dust-corrected FUV luminosity (Section \ref{subsec:cal_one}). We then also derive a ``multi--wavelength'' relation between the dust-corrected FUV luminosity and a linear combination of the observed (uncorrected for dust) FUV and 7.7~$\mu$m luminosities (Section \ref{subsec:cal_mult}).  In both cases, we evaluate the performance of our calibrations by comparing the estimated SFRs derived from both single and multi--wavelength SFR calibrations with SED model estimated SFRs from our work and the independent analysis from the \cite{Stefanon_2017} catalog. 
\par

\subsection{The PAH--FUV Relation}\label{subsec: pah_fuv}
\begin{figure*}[t]
\centering
\includegraphics[scale=0.6]{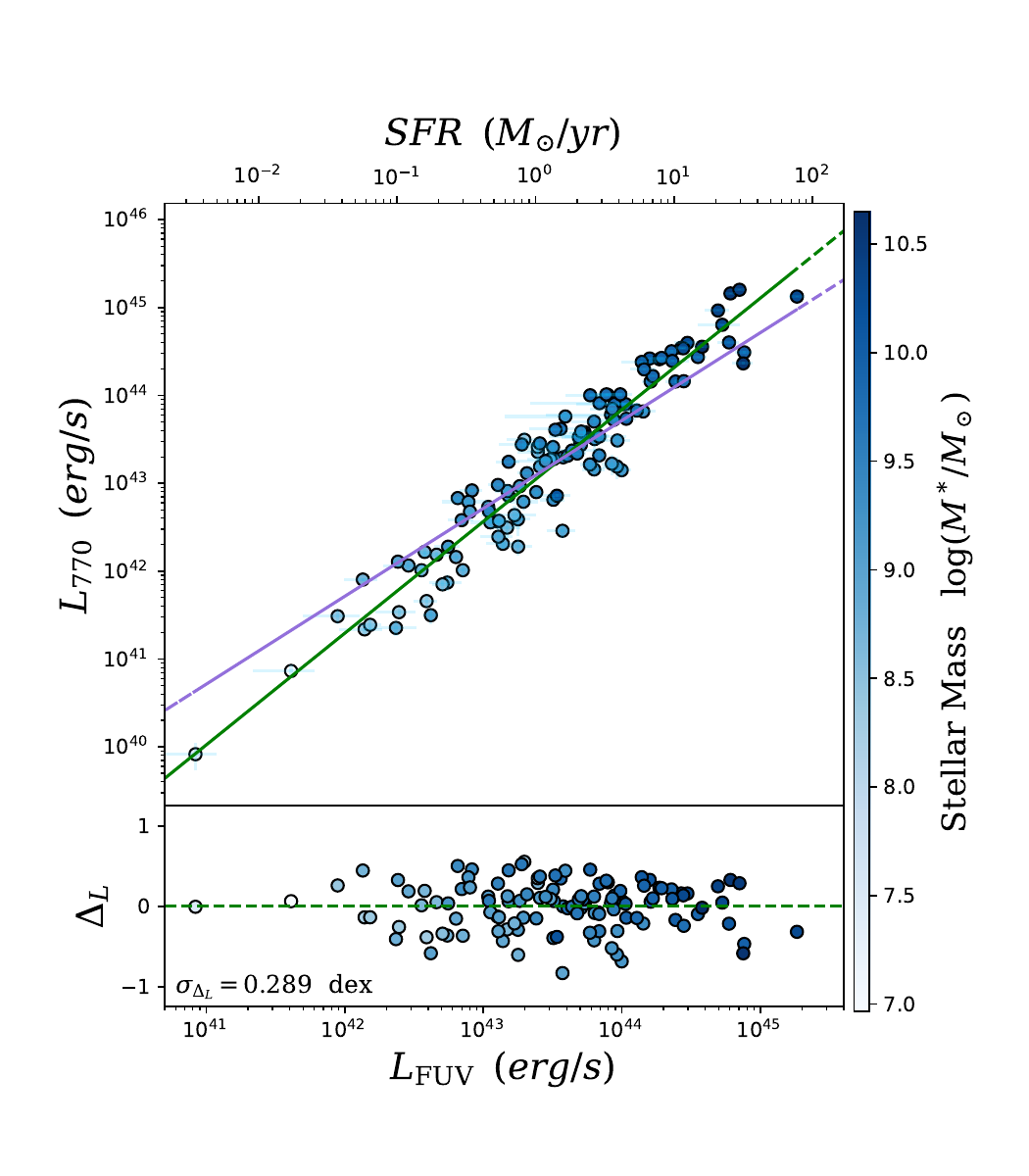}
\caption{(Top:) \(L_{770}\) compared to the dust-corrected FUV luminosity. The top axis shows the corresponding SFR derived following \cite{Kennicutt_2012} corrected for a \citealt{Chabrier2003} IMF. The green dashed line is the linear relationship (as described in Equation \ref{eqn:l770_linear}), and the purple dashed line shows the the unity relation (as described in equation \ref{eqn:l770_unity}). (Bottom:) Scatter about the linear fit where  \(\Delta L = \log(L_{770}) - \log(L_{Linear})  \), with $\sigma = 0.289$~dex. The data are colored by stellar mass in log scale. \label{fig:pahvsfuv}}
\end{figure*}

We compare the rest-frame dust-corrected FUV luminosity with the rest-frame 7.7 \(\mu m\) luminosity in Figure \ref{fig:pahvsfuv}. In this figure we introduce an additional top abscissa that shows the SFR corresponding to the dust-corrected FUV using the relation from \cite{Kennicutt_2012} assuming a constant SFR over the past 100 Myr, which is corrected for a \citealt{Chabrier2003} IMF using the conversion factor from \citealt{Madau_2014}. 

To characterize the correlation between the dust-corrected FUV and PAH luminosities, we fit a linear relation (where the slope is a free parameter) and unity 
relation (where the slope is set to one).  Specifically, we define the linear relation such that $\log L_{770} \propto k\times \log L_\mathrm{FUV}$, where $k$ is a constant of proportionality.  The unity relation then has $k=1$.  The linear fits were measured using \textsc{LINMIX}, a python package that uses a hierarchical Bayesian approach from \cite{Kelly_2007} and accounts for uncertainties on both the dependent and independent variables. From these we obtain, 
\begin{equation}\label{eqn:l770_linear}
    \begin{aligned}
        \log(L_{770}) = (1.27\pm0.04)\log(L_\mathrm{FUV}) - (12.1\pm2),
    \end{aligned}
\end{equation}
where the luminosities have units of erg~s$^{-1}$. 
The unity fits are measured using \emph{curve\_fit} from \textsc{scipy} \citep{Virtanen2020}, which gives  
\begin{equation}\label{eqn:l770_unity} 
    \begin{aligned}
        \log(L_{770}) = \log(L_\mathrm{FUV}) - (0.286     \pm0.001).
    \end{aligned}
\end{equation}
The fitted unity and linear relations are shown as the green and purple 
lines respectively in Figure~\ref{fig:pahvsfuv}. The deviation from unity at the luminosity limits of our sample suggests that for these regimes the PAH emission has a complex relation with the SFR. We further explore this in Section \ref{subsec:PAHcorrelationwithSFR}.\par

We test the unity and linear relations above using the  Akaike information criterion (AIC). The AIC considers an improvement in the likelihood of a the fit of a model with additional parameters, where a model is adopted if the change in the log-likelihood increases by more than the change in twice the number of parameters.   We find that the linear fit shows an improved log-likelihood by a factor of $\sim 1.4$ when we have added only one additional parameter. This indicates that the data favors the linear model over the unity model. For this reason we measure the scatter about the linear relation as opposed to the unity in the bottom panel of Figure \ref{fig:pahvsfuv}. Formally, we measure a scatter of 0.29 dex. \par 

\subsection{Single--Wavelength SFR Calibration}\label{subsec:cal_one}
Motivated by the strong correlation between the rest-frame mid-IR luminosity and the rest-frame dust-corrected FUV luminosity, we derive a ``single--wavelength’’ calibration of SFR from the 7.7$\,\mu$m luminosity. We convert $L_{770}$ to $L_{FUV}$ using the linear fit from equation~\ref{eqn:l770_linear}, which was selected by the AIC as mentioned in Section \ref{subsec: pah_fuv}. We then derive SFR from $L_{\mathrm{FUV}}$ following the relation from \cite{Kennicutt_2012} corrected for a \cite{Chabrier2003} IMF. The resulting conversion is,
\begin{equation}\label{eqn:sfr_cal_single}
\begin{aligned}
    \log(SFR_\mathrm{7.7 \mu m}) = (0.787\pm0.03)\log(L_{770} ) - (33.8\pm2),  
\end{aligned}
\end{equation}
where the SFR is measured in $M_\odot$~yr$^{-1}$ and the luminosity is again in units of erg~s$^{-1}$.\par

\begin{figure}[t]
\centering
\includegraphics[scale=0.45]{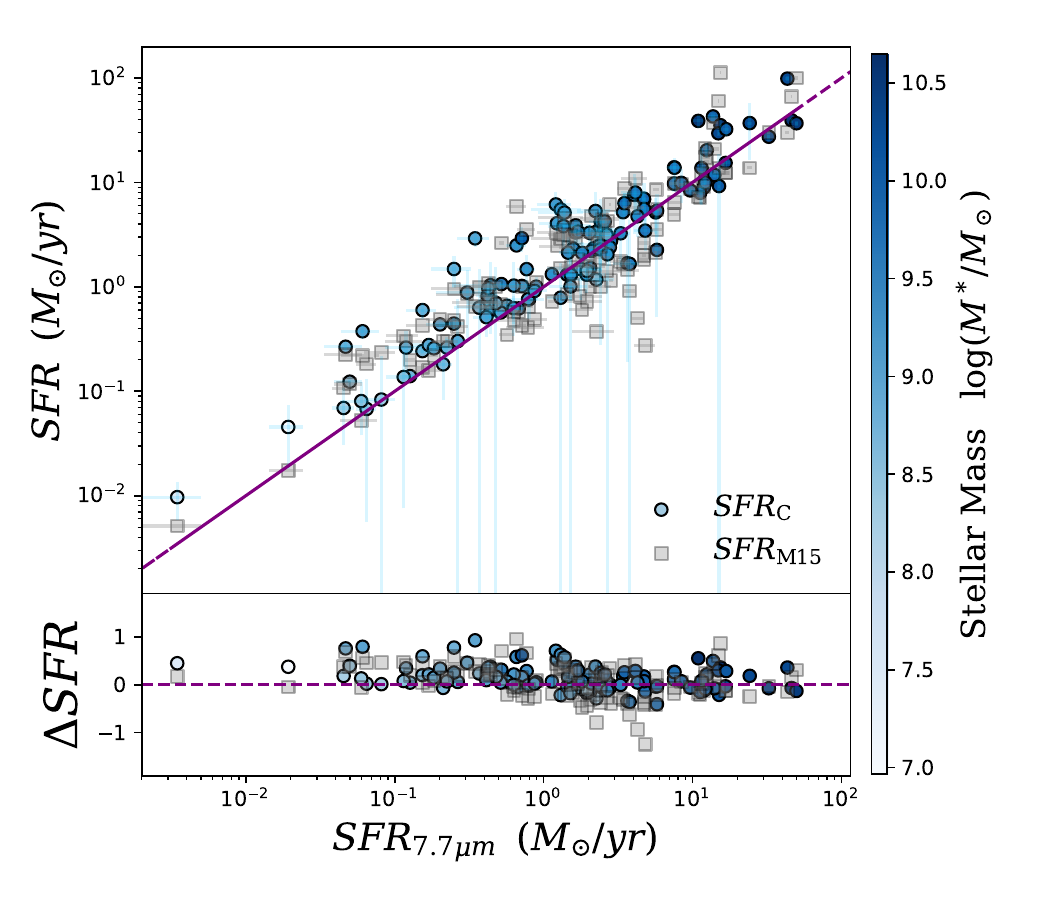}
\caption{(Top:) SFR derived from the single--wavelength calibration compared to SFR$_{\mathrm{C}}$ (the circular points colored by stellar mass estimated by \textsc{CIGALE}) and SFR$_{\mathrm{M15}}$ (represented with the grey squares). The dark purple line shows the one-to-one relation. (Bottom:) Measured scatter about the one-to-one relation, which is measured by \(\Delta SFR\) = \(\log(SFR_{7.7 \mu m})\) - \(\log(SFR)\). The measured $\sigma_{\Delta \mathrm{SFR}}$ for SFR$_{\mathrm{C}}$ and SFR$_{\mathrm{M15}}$  is 0.24 dex and 0.36 dex respectively.\label{fig:sfr_cal_unity} }
\end{figure}

To test the performance of this calibration we compare the estimated values for SFR$_\mathrm{7.7 \mu m}$ from Equation \ref{eqn:sfr_cal_single} to SFR$_\mathrm{C}$ and independently measured SFR estimates from the \cite{Stefanon_2017} catalog estimated with method 2a from \citep[][SFR$_{\mathrm{M15}}$]{Mobasher_2015} in Figure \ref{fig:sfr_cal_unity}. The work from \cite{Mobasher_2015} estimated SFR$_{\mathrm{M15}}$ by modeling the multi--wavelength photometry from the \cite{Stefanon_2017} catalog using a variety of methods. We adopt method ``2a'' from Mobasher et al, which was fixed to a \cite{Chabrier2003} IMF and left star formation history (SFH), metallicity, extinction, and population synthesis code as free parameters. We selected this SFR estimate from the \cite{Stefanon_2017} catalog since it was the most similar to our work. We measure the scatter between the model estimated SFRs to the estimates from our calibration to find $\sigma_{\Delta SFR}=$ 0.24 dex for SFR$_{\mathrm{C}}$ and 0.36 dex for SFR$_{\mathrm{M15}}$ respectively. We note that there appears to be an offset in the plot where the SFR$_C$ values are higher than SFR$_{7.7\mu m}$ at lower SFRs, but we expect this to be a result of the star-formation histories used by the SED modeling (see Appendix~\ref{sec:modelestimates}). Regardless, the measured scatter about the linear relation in Figure \ref{fig:sfr_cal_unity} is  tight across all 7.7~$\mu m$ luminosities.\par

\subsection{Multi--Wavelength SFR Calibration}\label{subsec:cal_mult}
We next consider a case where the total SFR is a combination of the unobscured SFR measured from the observed rest-frame FUV, and obscured SFR measured from the 7.7 $\mu$m luminosity. In principle, there should exist some energy-balance between these two variables as they trace the total emission from the young massive stars 
\citep[e.g.,][]{Calzetti_2007, Kennicutt_2007}. Motivated by this concept, we model the total intrinsic FUV luminosity as a multi-wavelength, linear combination of the observed FUV luminosity (\textit{un}-corrected for dust attenuation) and the 7.7 $\mu$m luminosity using , 
\begin{equation}\label{eqn:sfr_cal_multi}
    L_\mathrm{FUV} = L_{\mathrm{FUV}, obs} + \eta L_{770}.
\end{equation}

\noindent Equation \ref{eqn:sfr_cal_multi} is similar to the linear combination of L$_{\mathrm{TIR}}$ and the FUV luminosity discussed by Equation 11 from \cite{Kennicutt_2012}, and is in the simplest form.  In principle there could be additional factors that manifest as higher order polynomials, which we ignore here.  Using \emph{curve\_fit} from \textsc{scipy} we find that \(\eta = 0.732 \pm 0.002 \).\par

Using this result, we establish a ``multi--wavelength'' calibration for the SFR based on the linear combination of the observed FUV and the mid-IR luminosity using the FUV-SFR relation from \cite{Kennicutt_2012} corrected for a \cite{Chabrier2003} IMF. This yields, 
\begin{equation}\label{eqn:sfr_cal_multi_}
\log SFR_{\mathrm{FUV+7.7\mu m}}=  \log( L_{\mathrm{FUV}_{obs}} + \eta L_{770}) - 43.32, 
\end{equation}
where $\eta = 0.732\pm 0.002$ from above, the SFR is in units of $M_\odot$~yr$^{-1}$ and the luminosities are in units of erg~s$^{-1}$. \par 

\begin{figure}[b]
\centering
\includegraphics[scale=0.45]{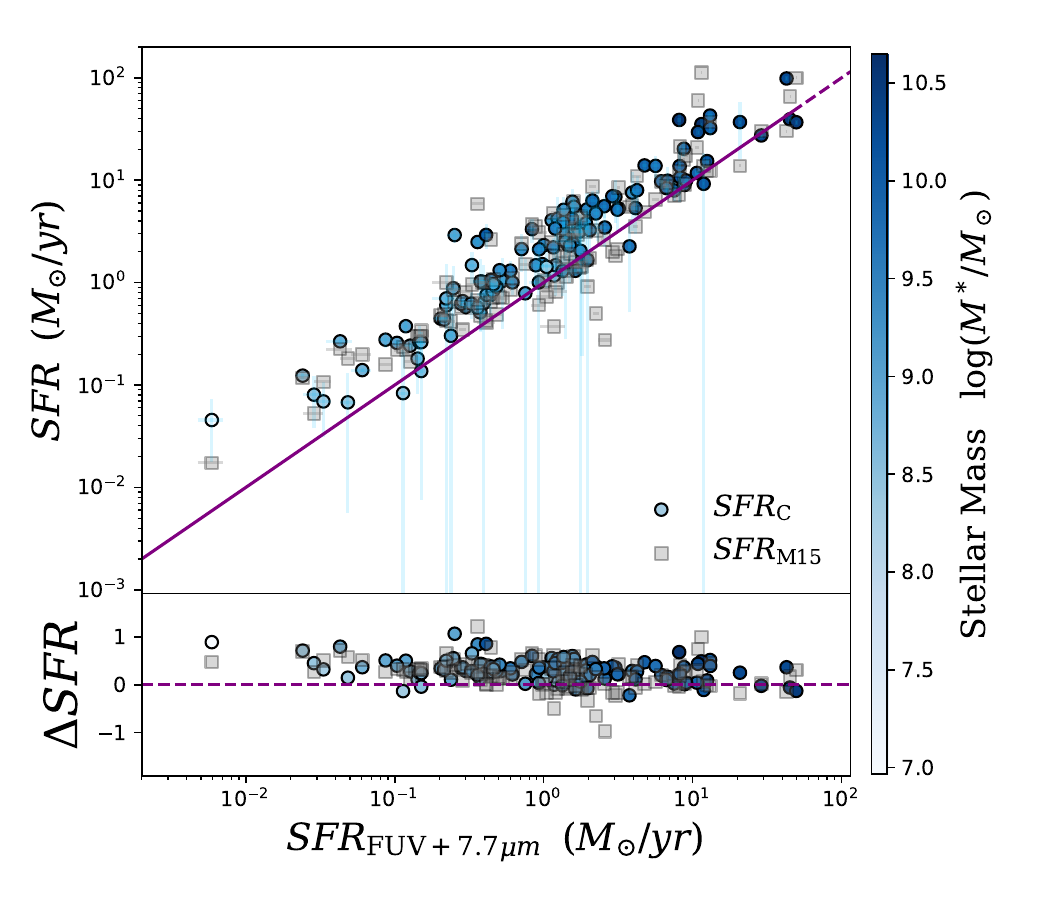}
\caption{(Top:) SFR derived from the multi--wavelength calibration (shown in Equation \ref{eqn:sfr_cal_multi_}) compared to SFR$_{\mathrm{C}}$ (represented with the circular points colored by stellar mass from \textsc{CIGALE}), and SFR$_{\mathrm{M15}}$ (represented with the grey squares). The dark purple line shows the one-to-one relation. (Bottom:) Measured scatter about the one-to-one relation, which is measured by \(\Delta SFR\) = \(\log(SFR_{\mathrm{FUV}})\) - \(\log(SFR)\). The measured $\sigma_{\Delta \mathrm{SFR}}$ for SFR$_{\mathrm{C}}$ and SFR$_{\mathrm{M15}}$  is 0.21 dex and 0.27 dex respectively.\label{fig:sfr_cal_multi}}
\end{figure}

We compare the estimated SFRs from the multi--wavelength calibration to the model estimated SFRs (SFR$_\mathrm{C}$ and SFR$_{\mathrm{M15}}$) in Figure \ref{fig:sfr_cal_multi}. We measure the scatter between the model estimated SFRs to that of the multi--wavelength calibration, which results in $\sigma_\mathrm{\Delta SFR}$ = 0.21 dex for SFR$_{\mathrm{C}}$ and 0.27 dex for SFR$_{\mathrm{M15}}$. We again observe an offset between the model estimated SFRs and the estimates from the multi--wavelength calibration at SFRs below roughly $10^{-1}\; M_\odot yr^{-1}$, which is attributable to the SFH used on the models in addition to varying galaxy properties. We explore the effects of SFH on SFR$_\mathrm{C}$ (see Appendix~\ref{sec:modelestimates}), and the galaxy properties which can contribute to the observed offset at these SFRs in Section \ref{subsubsec:PAHcorrelationwithSFR_low}.\par

\section{Discussion}\label{sec:discussion}
MIRI provides a new opportunity to quantify dust obscured star formation for high redshift galaxies that are much fainter than those previously accessible by any of the JWST predecessors. As we have demonstrated above, the rest-frame mid-IR luminosity measured from broad-band JWST/MIRI data ($L_{770}$) correlates strongly with the most recent star formation activity traced by the far-ultraviolet luminosity. We discuss the physical process behind the FUV-PAH correlation in Section~\ref{subsec:PAHcorrelationwithSFR}. We then discuss what types of galaxies depart from the FUV-PAH correlation in Sections~\ref{subsubsec:PAHcorrelationwithSFR_low} and \ref{subsubsec:PAHcorrelationwithSFR_high}. We place our results in the context of previous calibrations of the PAH luminosity in Section~\ref{subsec:lit_comparison}. Finally, we discuss some caveats that can impact the interpretation of our results in Section~\ref{subsec:caveats}.

\subsection{The Relation between the PAH Emission and SFR }\label{subsec:PAHcorrelationwithSFR}
In this section we consider the relation between the 7.7~$\mu m$ luminosity and the SFRs in three different regimes: galaxies with ``moderate'' SFRs ($\sim 10-30$~$M_\odot$~yr$^{-1}$) where the galaxies have PAH luminosities that are nearly proportional to the total SFR;  ``low'' SFRs ($\lesssim$ 10~$M_\odot$~yr$^{-1}$), where the PAH luminosities of the galaxies are low compared to the total SFRs; and ``high'' SFRs ($\gtrsim 30$~$M_\odot$~yr$^{-1}$), where the PAH luminosities again depart from the unity relation with FUV based SFRs as shown in Figure \ref{fig:pahvsfuv}. Each SFR regime is likely a result of different physical effects in galaxies that impact this relation. 

\subsubsection{Relation at Moderate SFRs}
Figure \ref{fig:pahvsfuv} highlights the capability of the PAH emission to trace the total SFR, where we compare the rest-frame 7.7~$\mu$m emission to the SFR from the dust-corrected FUV luminosity. We find that more than half of our sample (60\%) has SFRs between 10-30 $M_\odot$ yr$^{-1}$, which is where the linear and unity correlations between the dust-corrected FUV luminosity and the PAH luminosity intersect. Such galaxies provide important case-studies in which the 7.7 \(\mu m\) PAH luminosity is a direct tracer the total star formation rate.  This is consistent with previous studies that focused on the relation between PAH luminosity and the SFR (e.g., \citealt{Houck_2007,Shipley_2016}, see Section~\ref{subsec:lit_comparison} below). Here, the implication is that the 7.7~$\mu m$ luminosity is directly proportional to the SFR. The scatter in the relations is also small, with $\
\sigma_{\Delta\mathrm{SFR}} \simeq 0.3$~dex, which is likely a systematic floor to the (combined) accuracy of the UV and mid-IR SFRs. \par

To further investigate the reasoning behind this occurrence in our sample for this regime, we must first examine the properties of these sources. For this subset of our sample with SFR$\approx 10-30~M_\odot$ yr$^{-1}$, the galaxies are optically thick in the visual (the average dust attenuation is \(A_V \simeq 1.9\)) with an average stellar mass of 9.5 $\log(M^*/M_\odot)$.  This is typical of galaxies at these masses/SFRs, where  most of the star-formation in such galaxies is obscured.  For example, \citet{Whitaker2017} find that $\sim$ 70-90\% of star-formation is obscured for galaxies in this redshift and mass range. This is significant in the era of JWST as more obscured galaxies are being discovered due to the unprecedented sensitivity. It is also prudent to consider the galaxy properties in which such an assumption would not be valid, which we explore below.    \par 

\subsubsection{Relation at Low SFRs}\label{subsubsec:PAHcorrelationwithSFR_low}

From Figure \ref{fig:pahvsfuv} we observe that at low PAH luminosities (L$_{770}$ / erg~s$^{-1}$ $> 10^{42}$) the slope of the relation between the SFR and L$_{770}$ is steeper than the unity relation.  This means that the 7.7 \(\mu m\) PAH luminosity is weaker (at fixed SFR or at fixed mass), and is less of a direct tracer of the total SFR.  This occurs at stellar masses of approximately \(M < 10^8 M_\odot\).  This subset of our sample corresponds to the lowest redshift objects in our sample (\(z < 0.75\), see Figure \ref{fig:samplecompleteness}).\par

Observations of local galaxies show that the PAH emission is significantly weaker and less correlated with star 
formation for metal-poor galaxies.  In such objects the PAH emission drops by up to a factor of 30 for metal-poor \ion{H}{2} regions compared to metal--rich counterparts \citep{Engelbracht_2005,Calzetti_2007}. Due to the lack of necessary data to determine the metallicity content of our sample, we approximate the metallicity using a mass-metallicity relation (MZR) from \cite{Zahid_2011}. We find that these low mass sources have metallicities of $0.4$ dex below Solar, therefore we expect that the lower 7.7~\micron\ luminosities for low-mass galaxies is a result of lower metallicities.  This is consistent with previous work for nearby galaxies, as  seen in Figure 3 of \cite{Calzetti_2007} for \ion{H}{2} regions in  galaxies with intermediate and low metallicities ($12 + \log(\mathrm{O/H}) < 8.35$, i.e., less than about 0.5 dex of the Solar value). We note that the scatter in both of the SFR calibrations derived in this work are remarkably constant (it remains close to $\simeq$0.3 dex) even at low stellar masses. This likely implies there is a common cause to the decrease in the 7.7~$\mu m$ luminosity --- such as the galaxies having lower metallicity --- rather than some other mechanism that would lead to larger scatter. \par 

We consider other physical phenomena in galaxies besides low metallicity which could cause the PAH emission to not be capable of tracing the total SFR in these low mass galaxies. One alternative is that the lower PAH emission is caused by hard ionizing radiation fields that destroy the molecules, or delayed formation of PAH molecules in AGB stars \cite{Chastenet_2023}. Both of these require timescale arguments, we expect the galaxies to have a wider range of ages that allowed by the relatively low scatter between $L_{770}$ and SFR in our sample. Another possible explanation as to why the PAH luminosity does not directly trace the total SFR for lower-mass/SFR galaxies is that the degree of obscuration is lower in these galaxies. For three of the 14 low--mass sources we observe that these galaxies experience low attenuation, with (\(A_V \simeq 0.3\)).  As such, the dust and PAH molecules do not trace the majority of the light emitted by star-forming regions in galaxies \citep{Hirashita2001}. However, the majority of the low-mass sources (11/14) are more obscured (\(A_V > 1\)). This evidence suggests that lower metallicity is the more probable cause as to why the measured PAH luminosity is less correlated with the total SFR. \par

\subsubsection{Relation at High SFRs}\label{subsubsec:PAHcorrelationwithSFR_high}

Figure \ref{fig:pahvsfuv} shows that at high PAH luminosities ( L$_{770}$ / erg s$^{-1}$ $> 10^{44}$) the slope of the relation to the SFR is shallower than the unity relation, which indicates that the 7.7 \(\mu m\) PAH luminosity is departing as a direct tracer of the total SFR. Given the SFR, this occurs for galaxies with sellar masses above approximately \(10^{10.5} M_\odot\). Based on Figure \ref{fig:samplecompleteness}, these galaxies are between \(1.25 < z < 2\) and \(L_\mathrm{TIR} > 10^{11} L_\odot\).  However, we note again that the scatter in the relation between $L_{770}$ and the SFR remains relatively small, $\sigma_{\Delta\mathrm{SFR}} \simeq 0.3$~dex, which implies that the cause of this shallower relation between $L_{770}$ and the SFR is not a result in an increased scatter. \par

In these higher luminosity regimes, there are two primary reasons why the strength of the PAH features could be expected to diverge as a tracer of the total SFR.  One reason would be that the strength of PAH emission is suppressed in ultra-
luminous IR galaxies (ULIRGs) with $L_{\mathrm{TIR}} > 10^{12}$~$L_\odot$, as has been seen in some studies \citep[][]{Pope_2008, Takagi_2010, Rieke_2015}. A second reason is that these galaxies have built up a population of older, more-evolved stellar populations that are contributing to the heating of the PAH molecules, where it can be possible for both effects to contribute. We favor the second scenario as more applicable to our sample based on our discussion in the Appendix \ref{sec:binned_trends} (see also Figure~\ref{fig:pahtrends}), where we explore trends between the ratio of the dust-corrected FUV luminosity to the $L_{770}$ luminosity as a function of stellar mass and $A(\mathrm{FUV})$. We observe the $L_{770}$ luminosity is higher than the  dust-corrected FUV for galaxies at larger stellar mass, which indicates that there is an increase in the fraction of the PAH luminosity that is not correlated with the short-lived stellar populations that drive the UV emission. If instead there was an increase in the suppression of PAH molecules in ULIRG-type galaxies in our sample, we would anticipate the opposite outcome for higher stellar masses.  Again, both processes may be at play, but the lower scatter between $L_{770}$ and the SFR even for high SFR galaxies indicates most galaxies follow the same trends. One caveat here is that we excluded galaxies in the ``starburst'' phases that lie more than 0.6~dex above the star-forming main sequence, but could not identify starbursts that might lie within the star-forming main sequence. These galaxies could contribute to the observed trends or may show differing trends between $L_{770}$ and the total SFR, which we will explore in future work. \par

\subsection{Comparison to literature}\label{subsec:lit_comparison}

\begin{figure}[b]
\centering
\includegraphics[scale=0.5]{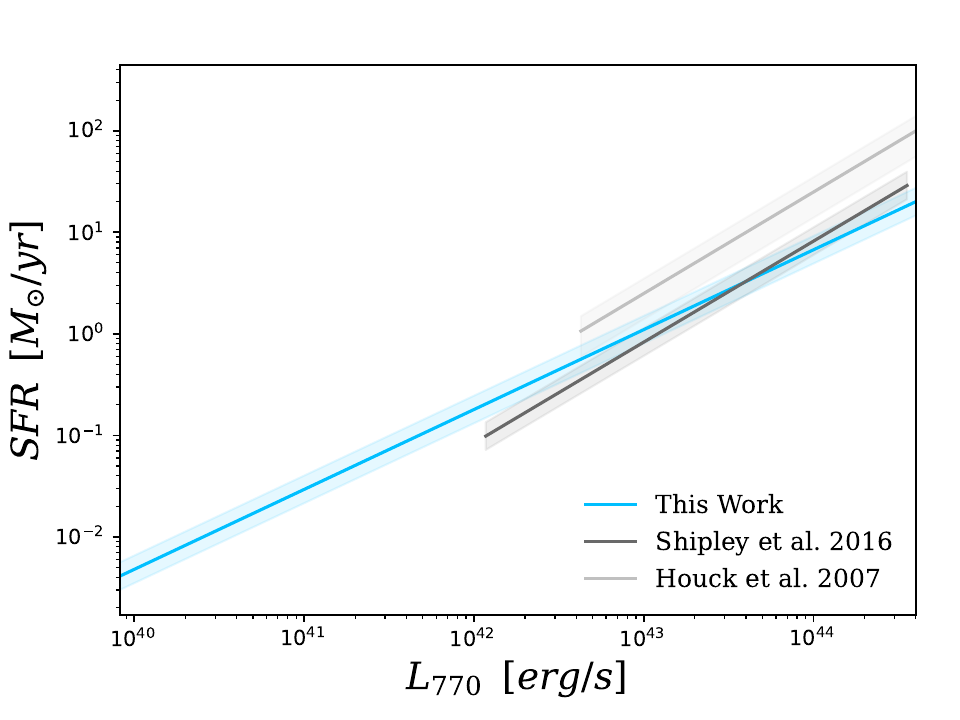}
\caption{Comparison of the $L_{770}$ and SFRs from our work and previous works in the literature. There is good agreement where the calibrations overlap, in particular between the relation from \citet{Shipley_2016} which used Spitzer/IRS data to predict the relation between the MIRI F770W rest-frame luminosity and SFR. All SFR estimates from the literature are corrected to an \cite{Chabrier2003} IMF. The shaded regions around each line are the 1 \(\sigma\) dispersion reported on each relation.\label{fig:lit_compare}}
\end{figure}

We compare our derived single--wavelength calibration to previous calibrations from literature in Figure \ref{fig:lit_compare}. To ensure an accurate comparison between calibrations, we corrected all calibrations to a \cite{Chabrier2003} IMF. There are other calibrations from literature \citep[][]{Hern_n_Caballero_2009, Xie_2019} that are not considered for our comparison due to the differences in sample selection, where these other studies include AGN and/or very high luminosity objects (e.g., bright ULIRGS) with a much larger SFR range that is not included in our sample. \par 

\cite{Shipley_2016} derived a relation between the PAH luminosities and the H$\alpha$ emission line, and used the relation between H$\alpha$ and SFR following \cite{Kennicutt_2012}. The 
sample of Shipley used Spitzer/IRS spectroscopy of relatively low-redshift galaxies, with \( z < 0.4\), where our sample probes much smaller SFRs at these same redshifts. To compare the calibration to our results, we select Equation 18 from \cite{Shipley_2016} and correct it to a 
\cite{Chabrier2003} IMF. The linear single--wavelength calibration was derived with a 
synthesized JWST/MIRI F770W filter, which was determined to be most similar to this 
work. We find that our calibration is consistent with the results from Shipley et al.\ within the 68\% uncertainties in the range where they are calibrated (see Figure~\ref{fig:lit_compare}).  \citet{Shipley_2016} concluded that the 7.7 $\mu m$ PAH feature directly traces the total SFR measured from 
dust--corrected H$\alpha$, with a unity relation.  Here, we find that the relation between 7.7 $\mu$m is linear, with a slope of $0.78$ (sub-unity).   The main reason for this difference is that we are considering galaxies over a larger range in luminosity, where we consider different effects that can impact the PAH emission (see Section~\ref{subsec:PAHcorrelationwithSFR}). In addition, \cite{Shipley_2016} used H$\alpha$-derived SFRs, which can probe the SFR on shorter timescales than the FUV. If we refit our linear relation derived in Equation \ref{eqn:l770_linear} to a subset of our sample at PAH luminosities which are comparable to Shipley et al.( $L_{770} \ge 1 \times 10^{43}$~erg s$^{-1}$), we would measure a slope of 1.11$\pm$ 0.07. This is consistent  with the single--wavelength calibration derived from Shipley et al., indicating that our ability to probe fainter SFRs with JWST/MIRI reveals the sub-unity relation at these lower luminosities. \par

\cite{Houck_2007} derived a relationship between the PAH luminosity and the total IR luminosity, and then used  the $L_{\mathrm{TIR}}$--SFR relation from \cite{Kennicutt_1998} to derive a single--wavelength SFR calibration for the PAH luminosity. The sample for their work spanned redshifts (\(z < 0.5\)) and included galaxies with a range of $L_{\mathrm{TIR}}$ and type, such as  AGN, ULIRGs, and starburst galaxies. To compare the results from Houck et al., to our calibration, we adjust the \cite{Kennicutt_1998} $L_{\mathrm{TIR}}$--SFR relation to the one from \cite{Kennicutt_2012} (accounting for the updated calibration and the  \cite{Chabrier2003} IMF). In general, the Houck et al.\ results return larger SFR values than the ones from both \cite{Shipley_2016} and this work.  This is evident in Figure~\ref{fig:lit_compare} as a small offset between the calibrations. The origin of this difference is likely related to the strength of the PAHs in different galaxies. For example, the PAH strength is observed to be weaker in AGN and ULIRGs \citep{Takagi_2010, Xie_2022}.  It could be possible that if one would want to calibrate SFRs using the PAH luminosities for these galaxies, then it would require larger area studies with JWST/MIRI to ensure proper statistics for high IR luminosity objects with well calibrated SFRs.  \par 

\subsection{Impact of Caveats and Assumptions}\label{subsec:caveats}
Estimates of parameters with \textsc{CIGALE} (e.g., SFRs, stellar mass, dust attenuation) can be less accurate for galaxies with high dust obscuration \citep{Pacifici2023}. This is one reason that we follow the recommendations of \cite{Pacifici2023} and use the measured properties from SED models that include FUV to Far-IR photometry.  Given that less that 1\% of our sources have any Hershel/PACS 100 or 160 $\mu m$ detections, the FUV attenuation is predominately constrained by the MIRI data. To test the effect of MIRI on FUV attenuation estimates from CIGALE, we reran the models from SED modeling Case 2 (as described in Section \ref{subsec:sedgen}) removing the MIRI data from the fits. We then compared the FUV attenuation estimates and found that the uncertainty in the FUV attenuation increases by an order of magnitude when the MIRI data are excluded. Both of our calibrations are dependent on dust-corrected FUV luminosity, which was corrected for dust with $A(\mathrm{FUV})$ output from \textsc{CIGALE}. This is why the MIRI data are included in our Case 2 SED modeling. We plan to study this further in a future work where we will  explore the mid-IR luminosity and SFR relation with SFR tracers that are less sensitive to attenuation compared to the FUV. \par 

We also consider if there is any stellar continuum contamination to $L_{770}$, which would cause our estimate to not strictly trace the 7.7 $\mu m$ PAH luminosity for our selected sample. To test this, we used the \textsc{CIGALE} model SED outputs that include the individual contributions from continuum, nebular, and dust spectra to the total SED (example 
shown in the top panel of Figure \ref{fig:MIRI_SEDs}). For sources with mid-IR 
luminosities above $L_{770} \gtrsim 1\times10^{43}$ erg~s$^{-1}$ the difference between the ``dust'' SED and the total SED is approximately 0.009 dex. Therefore, we conclude that the galaxies in our sample that are above these luminosities  do not have any contribution from the stellar continuum. For sources below $L_{770} \lesssim  
1\times10^{43}$ erg~s$^{-1}$, the PAHs are weaker, and the stellar continuum can account for some of the light. To quantify this, we examined the 14 sources below this luminosity limit and found that five appear to have small, but non-negligible  contribution of the stellar continuum to the $L_{770}$. The mean difference is approximately 0.06 dex between the ``dust'' SED model (that includes the PAH emission) and the the total SED from CIGALE for these five galaxies. Therefore, the PAH luminosity even in these cases accounts for than 87\% of the total mid-IR light.\par 

Lastly, we have assumed that the 7.7 $\mu m $ PAH luminosity can be reasonably measured by the average flux density in the rest-frame MIRI F770W bandpass.  Whereas, other studies have quantified the PAH luminosity as the integral of the emission in the lines directly. We test the validity of our assumption in Appendix \ref{sec:l770vsl77}, but will briefly describe it here. We used the estimated values of the 7.7 $\mu m $ luminosity from \cite{Kirkpatrick_2023} ($L_{7.7}$), which excludes the continuum emission and integrates over the emission of the line; this is described in Section 4.2 of their work. We then compared $L_{770}$ to $L_{7.7}$ in Figure \ref{fig:l770vsl77}.  We find there is a constant offset between the lines of 0.67 dex, and a tight scatter of 0.24~dex.  This offset is expected given the difference in the methods. In this work we compute $\nu \langle f_\nu \rangle$, in contrast to \citealt{Kirkpatrick_2014} who calculated the line flux as $F = \int F_\lambda\ d\lambda$. In Appendix \ref{sec:l770vsl77} we illustrate that this will lead to an offset of approximately $0.6$~dex, which accounts the near-constant offset (measured to be 0.67~dex) with a tight  scatter. \par

\section{Summary and Conclusions} 
\label{sec:summary_concl}
In this paper, we studied the relation between the mid-IR luminosity at 7.7 $\mu m$ and the SFR in star-forming galaxies at redshifts $0 < z < 2$.  We used photometry from CEERS MIRI, UVCANDELS, and the \cite{Stefanon_2017} multi--wavelength catalog and fit the SEDs with \textsc{CIGALE} for a sample of 120 galaxies. With the SED fits we measure the rest-frame FUV luminosity (uncorrected for dust attenuation) and the rest-frame 7.7 $\mu m$ luminosity from the average rest-frame flux in the MIRI F770W band (\(L_{770}\)). Using the best-fit estimates of the FUV attenuation ($A(\mathrm{FUV})$) from CIGALE, we correct the FUV luminosities for dust (\(L_\mathrm{FUV}\)). We then compare \(L_\mathrm{FUV}\) and \(L_{770}\), and from these we derive both a single--wavelength calibration between the SFR and $L_{770}$ and a multi--wavelength calibration between the SFR and a linear combination of the FUV luminosity (uncorrected for dust) and $L_{770}$. These calibrations are given in Equations \ref{eqn:sfr_cal_single} and \ref{eqn:sfr_cal_multi}. Our primary findings are as follows: 

\begin{itemize}[nosep, itemsep=5pt]
    \item We find that the 7.7 \(\mu m\) PAH luminosity is well correlated with 
        the dust-corrected FUV luminosity, following a linear relationship 
        described by Equation \ref{eqn:l770_linear}.
    \item Using the linear relationship between the 7.7 \(\mu m\) and dust-corrected FUV 
    luminosities, we derive a single--wavelength SFR calibration that approximates the total SFR with the obscured SFR in Equation \ref{eqn:sfr_cal_single}. We compare the SFR estimates from our single--wavelength calibration to model estimated SFRs from \textsc{CIGALE} and the SFRs from the independent catalog of \cite{Stefanon_2017}. The SFRs are well correlated with a  scatter of $\sigma_{\Delta \mathrm{SFR}}=$ 0.24 dex and 0.36 dex, respectively. We find that the total SFR can be approximated with the measured 7.7 $\mu m $ luminosity reliably for galaxies over a wide range of luminosity and dust attenuation.   
    \item We derive a multi--wavelength SFR calibration to estimate the the (dust-corrected) FUV based total SFR using a linear combination of the FUV luminosity (not corrected for dust) and the 7.7 \(\mu m\) luminosity. This method assumes an energy balance between the mid-IR and the FUV, which considers the total SFR as a combination of the unobscured and obscured SFR. We compare our SFR estimates from the multi-wavelength calibration to model estimated SFRs from \textsc{CIGALE} and the \cite{Stefanon_2017} catalog. From these we measure a scatter of $\sigma_{\Delta \mathrm{SFR}}=$ 0.21 dex and 0.27 dex, respectively.  The relatively small decrease in the scatter from the single-wavelength to the multi-wavelength calibration implies that these are near the systematic accuracy of the total SFR using either calibration. 
    \item We compare $L_{770}$ measured from the average flux in the rest-frame MIRI F770W bandpass to the independent estimate of the 7.7 \(\mu m\) luminosity ($L_{7.7}$) from \cite{Kirkpatrick_2023} and measure a scatter of 0.24 dex. Our estimates are offset from $L_{7.7}$ by 0.67 dex, which is primarily due to difference in the methods (this agrees with our estimate that the offset should be 0.6 dex based on the width of the MIRI F770W filter and various assumptions). This is further evidence that the mid-IR emission at 7.7 $\mu m$ is a good tracer of the SFR with a limiting systematic accuracy of approximately 0.2 -- 0.3~dex.   
\end{itemize}

This paper demonstrates the capability of the 7.7 \(\mu m\) PAH emission to trace star formation with JWST/MIRI. Future JWST surveys that explore the relation between the 7.7 \(\mu m\) feature and star formation in variable environments (such as starburts, ULIRGs, or AGN) will provide more insight into obscured star formation and the behavior of the 7.7 \(\mu m\) PAH feature in galaxies across redshifts. \par 
\software{astropy \citep{2013A&A...558A..33A,2018AJ....156..123A},  
          CIGALE \citep{Boquien_2019}, 
          Source Extractor \citep{1996A&AS..117..393B}, Matplotlib \citep{Hunter2007}, pandas \citep{jeff_reback_2022_6408044}, WebbPSF \citep{Perrin2012}, JWST Calibration Pipeline \citep{bushouse_howard_2022_7038885}, LINMIX \citep{Kelly_2007}
          }
\begin{acknowledgments}
This work benefited from support from NASA/ESA/CSA
James Webb Space Telescope through the Space Telescope
Science Institute, which is operated by the Association of
Universities for Research in Astronomy, Incorporated, under NASA contract NAS5-03127. Support for program No. JWST-ERS01345 was provided through a grant from the
STScI under NASA contract NAS5-03127. This work is based on observations with
the NASA/ESA Hubble Space Telescope obtained at the Space Telescope Science Institute, which is operated by the Association of Universities for Research in Astronomy, Incorporated, under NASA contract NAS5- 26555. Support for Program number HST-GO-15647 was provided through a grant from the STScI under NASA contract NAS5-26555. This work benefited from support from the George P. and Cynthia Woods Mitchell Institute for Fundamental Physics and Astronomy at Texas A\&M University. This research has made use of the Spanish Virtual Observatory (https://svo.cab.inta-csic.es) project funded by MCIN/AEI/10.13039/501100011033/ through grant PID2020-112949GB-I00. 
RAW acknowledges support from NASA JWST Interdisciplinary Scientist grants
NAG5-12460, NNX14AN10G and 80NSSC18K0200 from GSFC. 

\end{acknowledgments}

\begin{appendix} \label{sec:appendix}
\section{Model Estimated Star Formation Rates}\label{sec:modelestimates}
\textsc{CIGALE} calculates several different SFRs, including an instantaneous SFR, and the SFR averaged over 10 and 100 Myr timescales calculated from the star formation history (SFH). However, previous studies have shown that the SFR estimates can be biased because of the assumed parameterization of the SFH \citep[][]{Carnall_2019}. The FUV continuum is sensitive to the SFH over the past 100 Myr, and therefore one could expect that a SFR averaged over this timescale would be best correlated with the FUV--SFR relation from \cite{Kennicutt_2012}. This is only true if the SFH does not vary significantly over 100 Myr. For the case that the SFH varies on timescales faster than this, then the SFR/$L_\mathrm{UV}$ is time dependent and varies by factors of several to an order of magnitude \citep{Reddy_2012}. This is also true for SFHs that evolve exponentially in time (like those assumed here, see Table~\ref{tab:CIGALEparam}).  We therefore explore the impact of the SFH on the SED-measured SFRs here. \par

\begin{figure}[ht]
\centering
\includegraphics[scale=0.5]{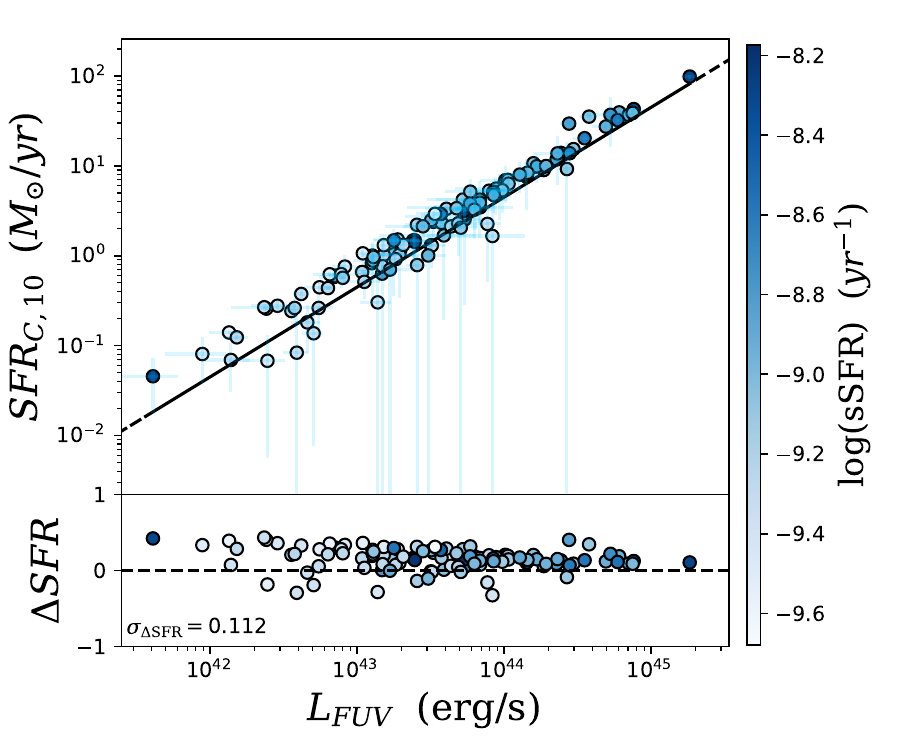}
\includegraphics[scale=0.5]{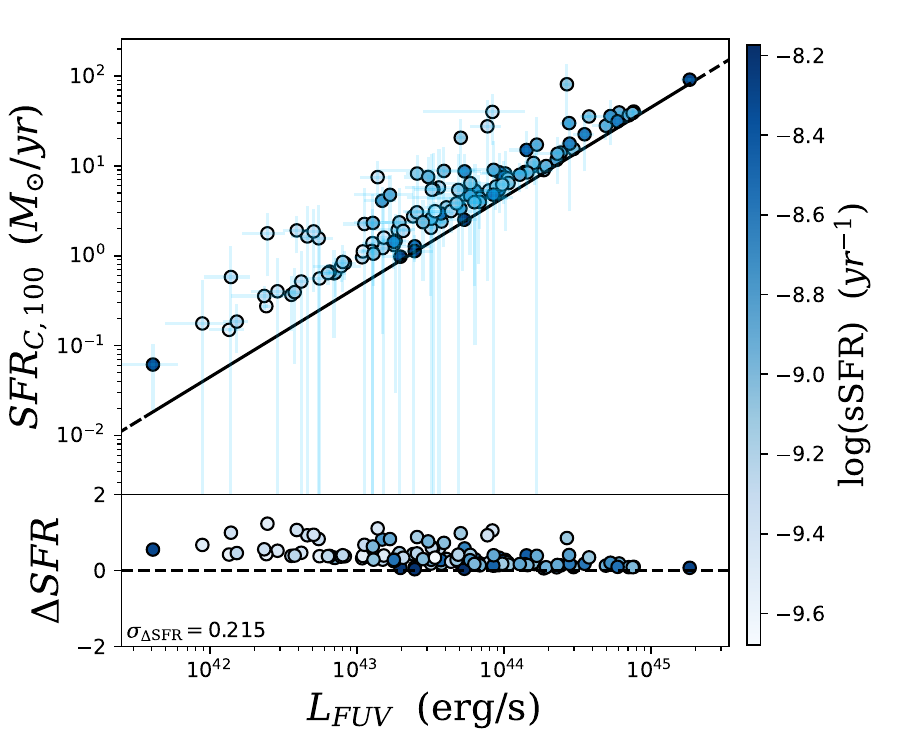}
\caption{(Left:) \textsc{CIGALE} output SFRs measured by the average SFR over 10 (Myr, \(SFR_{C,10}\)) from the SFH compared to dust-corrected FUV luminosity. (Right:) \textsc{CIGALE} output SFRs measured similarly to the left, but averaged over 100 (Myr, \(SFR_{C,100}\)). Both Figures are colored by sSFR in log scale and include the FUV-SFR relation from \cite{Kennicutt_2012} corrected to a \cite{Chabrier2003} IMF shown in black. For each panel we measure the scatter about the FUV-SFR relation using $\Delta SFR = \log(SFR_C,10/100) -\log(SFR)$}, where $\log(SFR)$ is the log of the FUV-SFR relation.
\label{fig:SFR_comparison}
\end{figure}

We compare the SFR averaged over 10 Myr (SFR$_{C,10}$) and those averaged of 100 Myr (SFR$_{C,100}$) to the dust-corrected FUV luminosity in Figure \ref{fig:SFR_comparison}. These plots show that while there is a strong correlation, the plots diverge from the FUV-SFR relation from \cite{Kennicutt_2012} at lower SFRs. Ultimately, we find that SFR$_{C,10}$ shows a tighter relation to the FUV-SFR relation with a measured scatter of 0.11 dex. In contrast, the SFR$_{C,100}$ values show a larger scatter of 0.215 dex.  We therefore use the SFR$_C$ values from \textsc{CIGALE} derived by averaging the SFH over the past 10~Myr. We do note, however, that there is an offset between the \textsc{CIGALE} SFR$_C$ values and the $L_\mathrm{FUV}$ values at lower SFRs. 
We interpret this as a result of uncertainties in the assumed SFHs.  This offset leads to the offsets seen in our relations in the main text (Figures~\ref{fig:sfr_cal_unity} and \ref{fig:sfr_cal_multi}), which we again attribute to the SFHs from \textsc{CIGALE}.   

\section{The relation between the 7.7~$\mu m$ Luminosity and the PAH luminosity}\label{sec:l770vsl77}

In this work we use the rest-frame mid-IR luminosity measured in the MIRI F770W bandpass. This bandpass includes the emission from the 7.7~$\mu$m PAH feature, which is the primary feature we use as a tracer of the SFR.   However, the F770W bandpass includes the 7.7 $\mu m$ PAH complex, the mid-IR continuum, and for some sources the Ar[II] and 8.6 $\mu m$ PAH feature \citep[][]{Pagomenos_2018}. While we expect the 7.7~$\mu$m emission to dominate the total emission in this band based on observations of local star-forming galaxies \citep[][]{Chastenet_2023}, here we consider how much of the rest-frame F770W luminosity stems from the PAH emission. \par 

\begin{figure}[ht]
\centering
\includegraphics[scale=0.5]{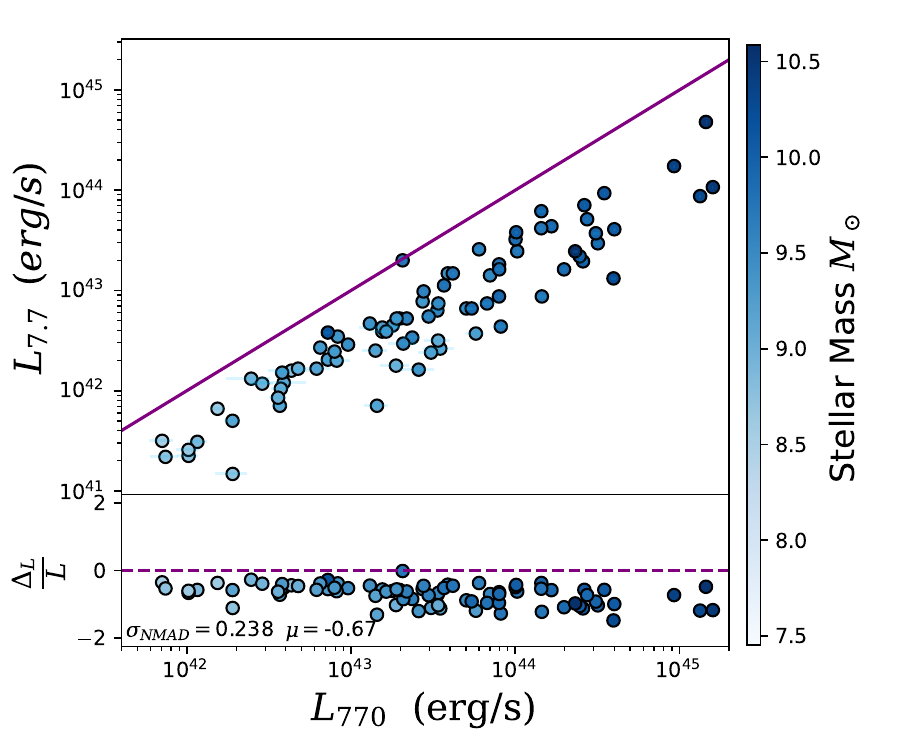}
\caption{(Top:) The 7.7 $\mu m$  luminosity estimated from \cite{Kirkpatrick_2023} compared to $L_{770}$, with points colored by stellar mass estimated from \textsc{CIGALE} in log scale. The dark purple line shows the one to one relation. (Bottom:) Measured scatter about the one-to-one relation, which is measured by $\Delta L = \log(L_{770}) - \log(L_{7.7})$. The data are colored by stellar mass in log scale. We argue that most of this offset is expected from the differences in the different methods (see text). 
 \label{fig:l770vsl77}}
\end{figure}

We compare our measurements of the PAH luminosity measured from the average flux in the rest-frame MIRI F770W bandpass ($L_{770}$) to the estimated 7.7 $\mu m$ luminosity ($L_{7.7}$) from \cite{Kirkpatrick_2023} in Figure \ref{fig:l770vsl77}.  Kirkpatrick et al.\ independently measured the luminosity in the 7.7 $\mu m$ PAH complex for galaxies in CEERS.  Here, we cross-correlated the galaxies in our sample with those from Kirkpatrick et al., finding 86 galaxies in common to both samples. Figure \ref{fig:l770vsl77} compares the mid-IR luminosities from our work ($L_{770}$) with the 7.7 $\mu$m PAH luminosities estimated by Kirkpatrick et al. A full description of the estimation of $L_{7.7}$ from Kirkpatrick et al. can be found in Section 4.2 of their work (the method is the same as the measurement for $L_{6.2}$ in Section 4.2), but we will briefly describe it here. This work uses mid-IR spectroscopy to estimate the continuum contribution to the 7.7\,$\mu$m feature with the 5MUSES sample that was observed with Spitzer/IRS. Kirkpatrick et al. selected 11 star-forming galaxies from the 5MUSES sample, which span redshifts 0.06-0.24 and $\log L_{\rm IR}=10.79-11.63\,L_\odot$. Kirkpatrick et al. shifted the 5MUSES spectra into rest-frame for sources that covers the 7.7\,$\mu$m feature. They then calculated $L_{7.7}$(5MUSES) by fitting a line to the continuum at 7.2 and 8.2\,$\mu$ to remove the continuum and then integrating the remaining luminosity. Kirkpatrick et al. also calculates a synthetic photometric point, $L_\nu$, by convolving with the appropriate transmission curve. They used the ratio $L_{7.7}/L_\nu$ for the 5MUSES galaxies to estimate $L_{7.7}$ for the MIRI galaxies, which we use for this work. \par

We find that $L_{770}$ is greater than $L_{7.7}$ roughly by 0.67 dex with a measured scatter of 0.24 dex. 
The reason for this offset is in likely because of the difference in the methods. Here we take the average flux density in the rest-frame F770W bandpass, $\nu \langle f_\nu \rangle$, while \citet{Kirkpatrick_2014} integrate over the continuum-subtracted line to get the total line flux, $F$.    Assuming the continuum is negligible (see above), we can take the line flux to be $F = \langle f_\lambda \rangle \Delta \lambda$, where $\langle f_\lambda \rangle$ is the average flux density, and $\Delta \lambda$ is the width of the F770W bandpass ($\Delta \lambda = 1.95$~$\mu m$).  We further set $\nu \langle f_\nu \rangle = \lambda \langle f_\lambda \rangle$ and calculate the ratio.  This leads to $\nu \langle f_\lambda \rangle / F \approx \lambda / \Delta \lambda = 7.7~\mu m/ 1.95~\mu m$, which is approximately a factor of 4, or (in logarithmic units) 0.6~dex. This is very nearly the observed offset (0.67~dex) and  therefore reasonably accounts for the near constant offset and tight scatter between the methods.

\section{Binned PAH Luminosity Trends}\label{sec:binned_trends}
We test if there is any dependence between the dust-corrected FUV luminosity and the 7.7~$\mu m$ luminosity as a function of galaxy stellar mass and FUV attenuation ($A(\mathrm{FUV})$), using the values estimated by \textsc{CIGALE}. If rest-frame 7.7~$\mu m$ luminosity directly traces the total SFR, we expect this ratio to be \(\simeq 1\). \par
\begin{figure}[ht]
\centering
\includegraphics[scale=0.5]{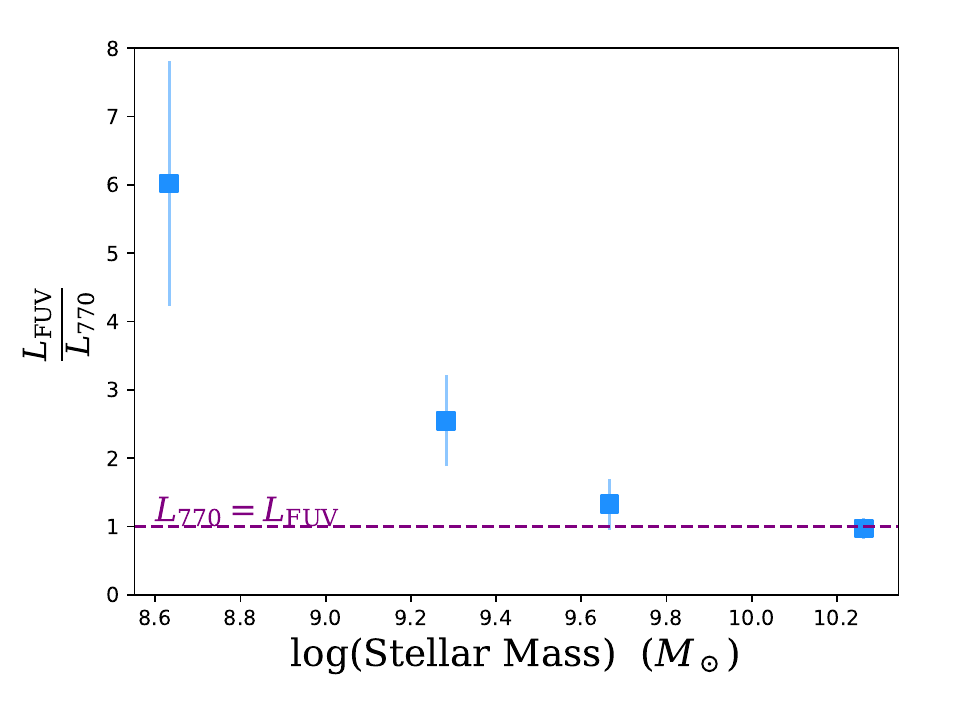}
\includegraphics[scale=0.5]{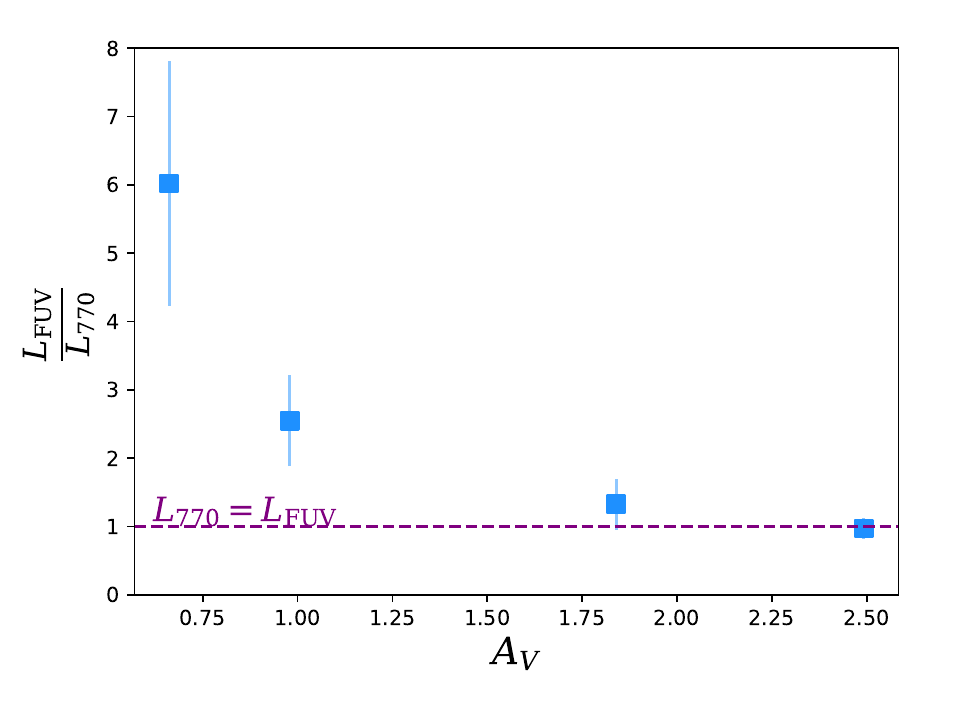}

\caption{Binned ratio between dust-corrected FUV and PAH luminosity compared to the stellar mass in log scale (left) and FUV attenuation ($A(\mathrm{FUV})$) (right) . Both the stellar mass and $A(\mathrm{FUV})$ are estimated from \textsc{CIGALE} as described in Section \ref{subsec:sedgen}. The dark purple dashed line in both panels are the \(L_\mathrm{FUV} = L_{770}\) line. \label{fig:pahtrends}}
\end{figure}

 
To study any general trends in the data we measure ratio of the  dust-corrected UV luminosity to the 7.7~$\mu m$ luminosity as a function of stellar mass and $A(\mathrm{FUV})$. Figure \ref{fig:pahtrends} shows the results.  At lower stellar masses (\(< 10^{9.4} M _\odot\)) the galaxies in our sample have much higher ratios of FUV luminosity to the 7.7 $\mu m$ luminosity, which reaches as high as a factor of 6. These galaxies tend to be optically thin ($A_V < 1$). For galaxies with higher stellar masses (\(> 10^{9.4} M_\odot\)) the dust attenuation and the 7.7~$\mu m$ luminosity increase with increasing stellar mass. In this case, we have already shown that the 7.7~$\mu m$  luminosity scales with the total SFR (see Figure \ref{fig:pahvsfuv}). The ratio of $L_\mathrm{FUV}/L_{770}$ dropping to unity for high stellar masses and high dust attenuation implies the existence of an additional source of PAH heating in these galaxies. This is expected as there exists heating from longer-lived stellar populations instead of from \ion{H}{2} regions \citep[][]{Boselli_2004}.  Therefore, it is most likely that these observed trends in our sample are caused by additional PAH heating (most likely from older stars) for galaxies at such high stellar mass. \par

\end{appendix}

\bibliography{citedpapers}{}
\bibliographystyle{aasjournal}
\end{document}